\documentclass[prb,aps,twocolumn,showpacs]{revtex4-1}

\usepackage{graphicx}
\usepackage{latexsym}
\usepackage{amsmath}
\usepackage{bm}

\newcommand{\be}{\begin{eqnarray}}
\newcommand{\ee}{\end{eqnarray}}

\begin{document}

\title{Effective field theories for two-component repulsive bosons on lattice \\
and their phase diagrams
}

\date{\today}

\author{Yoshihito Kuno} 
\author{Keisuke Kataoka} 
\author{Ikuo Ichinose}
\affiliation{%
Department of Applied Physics, Nagoya Institute of Technology,
Nagoya, 466-8555 Japan}
%{\today}

\begin{abstract}
In this paper, we consider the bosonic
t-J model, which describes two-component hard-core bosons with 
a nearest-neighbor (NN) pseudo-spin interaction and a NN hopping.
To study phase diagram of this model, we derive effective field theories 
for low-energy excitations.
In order to represent the hard-core nature of bosons, we employ a 
slave-particle representation.
In the path-integral quantization, we first integrate our the radial 
degrees of freedom of each boson field and obtain the low-energy
effective field theory of phase degrees of freedom of each boson field
and an easy-plane pseudo-spin.
Coherent condensates of the phases describe, e.g., a ``magnetic order"
of the pseudo-spin, superfluidity of hard-core bosons, etc.
This effective field theory is a kind of extended quantum XY model, 
and its phase diagram can be investigated precisely 
by means of the Monte-Carlo simulations.
We then apply a kind of Hubbard-Stratonovich transformation to the quantum XY model
and obtain the second-version of the effective field theory, which
is composed of fields describing the pseudo-spin degrees of freedom
and boson fields of the original two-component hard-core bosons.
As application of the effective-field theory approach, we consider 
the bosonic t-J model on the square lattice and also on the triangular lattice,
and compare the obtained phase diagrams with the results of the numerical
studies.
We also study low-energy excitations rather in detail in the effective field theory.
Finally we consider the bosonic t-J model on a stacked triangular lattice
and obtain its phase diagram.
We compare the obtained phase diagram with that of the effective field theory
to find close resemblance.
\end{abstract}

\pacs{67.85.Hj, 75.10.-b, 03.75.Nt}

\maketitle
%%%%%%%%%%%%%%%%%%%%%%%%%%%%%%
\section{Introduction}

In recent years, systems of cold atoms have attracted interest of many
condensed-matter physicists.
In particular, cold atoms put on an optical lattice sometimes regarded
as a ``simulator" to study canonical models of strongly correlated
electron systems\cite{optical,lw}.
These systems are quite controllable and impurity effects are suppressed.
Systems of single-species bosonic cold atoms are described by the
standard Bose-Hubbard model\cite{BHM}, which explains the Mott-superfluid 
phase transition.
For two-component bosonic systems, even the Mott phase exhibits rich
behavior\cite{exchange}.
At total filling factor one and with strong inter and intra-species repulsive
interactions, the two-component Bose-Hubbard model reduces to an
effective spin model, i.e. Heisenberg spin model\cite{SpinM},
which has a nontrivial phase diagram.
The experimental realization of bosonic mixture of $^{87}$Rb -$^{41}$K
in a three-dimensional optical lattice is an important development\cite{RbK}.
The two-component Bose-Hubbard model at commensurate fillings
has been studied in e.g., Refs.\cite{altman,sansone}
by the mean-field-theory (MFT) type approximations and the numerical methods, 
and its one-dimensional
counterpart by the Tomonaga-Luttinger liquid theory in Ref.\cite{Hu}.
Phase diagrams of these systems were clarified.

From the view point of the strongly-correlated condensed matter physics,
it is very interesting to study the two-component Bose-Hubbard model at 
{\em fractional fillings and with strong repulsions}.
Then, we have started study of the bosonic t-J model\cite{BtJ,BtJ1}, 
which is an effective model of the two-component Bose-Hubbard model at the
above limit and is also a kind of bosonic counterpart of the t-J model for 
the high-$T_c$ superconducting phenomena\cite{BtJ2}.
The bosonic t-J model (B-t-J model) describes two-component bosonic cold atoms
in an optical lattice with strong inter and intra-species repulsions that can be
controlled in the experiments.

Our previous studies on the B-t-J model mostly employed numerical Monte-Carlo (MC)
simulations to clarify its phase diagram at finite temperature ($T$)\cite{BtJ}.
In contrast to the fermionic t-J model, the numerical study can be done
without any difficulties for some cases of the B-t-J model.
In this paper, we shall derive effective fields theories of the B-t-J model
and study them by both analytical and numerical methods.
We mostly focus on the B-t-J model on square and triangular lattices
at $T=0$ and the phase diagram for quantum phase transitions.
The obtained results are compared with the previous findings and 
numerical study of finite-$T$ systems.

This paper is organized as follows.
In Sec.II, we shall introduce the B-t-J model and derive first-version of
an effective field theory by using the path-integral methods.
The local constraint is faithfully treated by using the slave-particle representation.
By integrating out amplitude modes of the slave particles, we obtained an extended
quantum XY model, which describe Bose condensation of two atoms and
pseudo-spin degrees of freedom.
Numerical study by the Monte-Carlo (MC) methods can be performed without
any difficulties and the phase diagram is obtained for the B-t-J model 
on both the square and triangular lattices in Sec.III.
In Sec.IV, we shall derive second-version of the effective field theory from the
extended quantum XY model by introducing collective fields for the two bosons
and the pseudo-spin degrees of freedom.
By using this effective field theory, we obtain phase diagram of the B-t-J model
and verify the consistency of the results.
Furthermore we study low-energy excitations, i.e., the Nambu-Goldstone bosons
in various phases in the phase diagram.
In Sec.V, we numerically shall the B-t-J model in the stacked triangular
lattice at finite $T$.
Obtained phase diagram has a similar structure of the model in the triangular lattice
at $T=0$.
Section VI is devoted for conclusion and discussion.

%%%%%%%%%%%%%%%%%%%%%%%%%%%%%%
\setcounter{equation}{0}
\section{Bosonic $\mbox{t-J}$ Model and derivation of effective model}

Hamiltonian of the B-t-J model, which will be studied in this paper, 
is given as\cite{BtJ,BtJ1,BtJ2,SpinM},
\begin{eqnarray}
H_{\rm tJ}&=&-\sum_{\langle i,j\rangle} t(a^\dagger_{i}a_j
+b^\dagger_{i}b_j+\mbox{h.c.})
+J_z\sum_{\langle i,j\rangle}S^z_{i}S^z_j  \nonumber  \\
&& +J\sum_{\langle i,j\rangle}(S^x_{i}S^x_j+S^y_{i}S^y_j)
%-{\mu}_c\sum_{r}(1-n_{ar}-n_{br}),
\label{HtJ}
\end{eqnarray}
where $a^\dagger_i$ and $b^\dagger_i$ are 
boson creation operators\cite{HCboson} at site $i$,
pseudo-spin operator $\vec{S}_i={1 \over 2}B^\dagger_i\vec{\sigma}B_i$ with
$B_i=(a_i,b_i)^t$, $\vec{\sigma}$ are the Pauli spin matrices,
and  $\langle i,j\rangle$ denotes nearest-neighbor (NN) sites of the lattice.
We shall consider both the square and triangular lattices in the following study.
Physical Hilbert space of the system consists of states with total particle
number at each site less than unity (the local constraint).
In order to incorporate the local constraint faithfully, we use the following
slave-particle representation\cite{BtJ,BtJ1},
\begin{eqnarray}
&& a_i=\phi^\dagger_i \varphi_{i1}, \;\;\; 
b_i=\phi^\dagger_i \varphi_{i2},  \label{slave}  \\
&& \Big(\phi^\dagger_i\phi_i+\varphi^\dagger_{i1}\varphi_{i1}+
\varphi^\dagger_{i2}\varphi_{i2}-1\Big)
|\mbox{phys}\rangle =0,
\label{const}
\end{eqnarray}
where $\phi_i$ is a boson operator that {\em annihilates hole} at site $i$,
whereas $\varphi_{1i}$ and $\varphi_{2i}$ are bosons that represent the pseudo-spin 
degrees of freedom.
$|\mbox{phys}\rangle$ is the physical state of the slave-particle Hilbert space. 

The previous numerical study of the B-t-J model\cite{BtJ,BtJ1,KKI} 
show that there appear 
various phases including superfluid with Bose condensation, state with
the pseudo-spin long-range order, etc.
For the most of them, the MC simulations show that density fluctuation 
at each lattice site is not large even in the spatially inhomogeneous states
like a phase-separated state.
From this observation, we expect that there appears the following term effectively,
\begin{eqnarray}
H_{\rm V} &=& {V_0 \over 4}\sum_i\Big(
(\varphi^\dagger_{1i}\varphi_{1i}-\rho_{1i})^2+
(\varphi^\dagger_{2i}\varphi_{2i}-\rho_{2i})^2 \nonumber \\
&& \hspace{2cm} +(\phi^\dagger_{i}\phi_i-\rho_{3i})^2
\Big),
\label{HV}
\end{eqnarray}
where $\rho_{1i}$ etc are the parameter that controls the densities of
$a$-atom and $b$-atom at site $i$, and $V_0(>0)$ controls their
fluctuations around the mean values.
It should be remarked here that the expectation value of the particle
numbers in the physical state $|\mbox{phys}\rangle$ are given as
${1 \over N}\sum_i \langle a^\dagger_i a_i \rangle\equiv
{1 \over N}\sum_i\mbox{Tr}_{\rm phys}(a^\dagger_ia_i)
={1 \over N}\sum_i\mbox{Tr}_{\rm phys}(\varphi^\dagger_{1i}\varphi_{1i})$
and similarly 
${1 \over N}\sum_i \langle b^\dagger_i b_i \rangle=
{1 \over N}\sum_i\mbox{Tr}_{\rm phys}(\varphi^\dagger_{2i}\varphi_{2i})$,
where $N$ is the number of sites of the lattice and $\mbox{Tr}_{\rm phys}$
denote the trace over the states satisfying the local 
constraint, i.e., the physical-state condition (\ref{const}).
Therefore
the constraint (\ref{const}) requires $\sum_{\sigma=1}^3\rho_{\sigma i}=1$ at each
site $i$.
The values of $V_0$ and $\rho_{\sigma i} (\sigma=1,2,3)$ are to be determined 
in principle by $t, \ J_z, \ J$ and 
filling factor, but here we add $H_{\rm V}$ to $H_{\rm tJ}$ by hand
and regard parameters in  $H_{\rm V}$ as a free parameter.
%After obtaining physical result, we shall check the consistency of the
%parameter assignment.
In this sense, we are considering an extended bosonic t-J model.
It should be stressed here that the strong on-site repulsions between
atoms enhance the effects of spatial lattice and as a result quasi-excitations
with Lorentz-invariant dispersion can appear in a non-relativistic 
original model like the Bose-Hubbard model\cite{Lorentz}.

The existence of $H_{\rm V}$ in Eq.(\ref{HV}) is very useful for study of the quantum
many-particle systems.
In the path-integral representation of the partition function $Z$, the action contains
the imaginary terms like $\int d\tau\bar{\phi}_i(\tau)\partial_\tau \phi_i(\tau)$,
where $\bar{\phi}_i$ is for $\phi^\dagger_i$ and $\tau$ is the imaginary time, i.e.,
\begin{eqnarray}
Z&=& \int [D\phi D\varphi_1 D\varphi_2]
\exp\Big[-\int d\tau\Big(\bar{\varphi}_{1i}(\tau)\partial_\tau \varphi_{1i}(\tau)
\nonumber \\
&& \hspace{1cm} +\bar{\varphi}_{2i}(\tau)\partial_\tau \varphi_{2i}(\tau)
 +\bar{\phi}_i(\tau)\partial_\tau \phi_i(\tau)  \nonumber \\
&& \hspace{1cm} +H_{\rm tJ}+H_{\rm V}
\Big)\Big],
\label{Z}
\end{eqnarray}
where $H_{\rm tJ}$ is expressed by the slave particles and 
the above path integral is calculated under the constraint (\ref{const}).
(In this paper we set $\hbar=1$.)
For the existence $H_{\rm V}$, we separate the path-integral variables 
$\varphi$'s and $\phi$ as 
\begin{eqnarray}
&&\varphi_{1i}=\sqrt{\rho_{1i}+\ell_{1i}}\exp(i\omega_{1i}), \nonumber \\  
&&\varphi_{2i}=\sqrt{\rho_{2i}+\ell_{2i}}\exp(i\omega_{2i}), \label{param1} \\
&&\phi_i=\sqrt{\rho_{3i}+\ell_{3i}}\exp(i\omega_{3i}), \nonumber
\end{eqnarray}
and we integrate out the (fluctuation of)
radial degrees of freedom.
There exists a constraint like 
$\ell_{1i}+\ell_{2i}+\ell_{3i}=0$
on performing the path-integral over the radial degrees of freedom, i.e.,
$\ell_{\sigma i} \ (\sigma=1,2,3)$.
This constraint can be readily incorporated by using a Lagrange 
multiplier $\lambda_i(\tau)$,
$$
\prod_\tau\delta(\ell_{1i}+\ell_{2i}+\ell_{3i})=\int d\lambda_i
e^{i\int d\tau (\ell_{1i}+\ell_{2i}+\ell_{3i})\lambda_i}.
$$
The variables $\ell_{\sigma i} \ (\sigma=1,2,3)$ also appear in $H_{\rm tJ}$,
but we ignore them by simply replacing 
$\varphi_{\sigma i} \rightarrow \sqrt{\rho_{\sigma i}}\exp(i\omega_{\sigma i})$,
and then we have
\begin{eqnarray}
&&\int d\lambda_i d\ell_{i}e^{\int d\tau\sum_{\sigma=1}^3(-V_0(\ell_{\sigma,i})^2
+i\ell_{\sigma,i}(\partial_\tau \omega_{\sigma,i}+\lambda_i))} \nonumber   \\
&& \hspace{1cm} =\int d\lambda_i e^{-{1 \over 4V_0}\int d\tau\sum_\sigma
(\partial_\tau \omega_{\sigma,i}+\lambda_i)^2},
\label{integral}
\end{eqnarray}
where we have ignored the terms like $\int d\tau \partial_\tau \omega_{\sigma, i}$
by the periodic boundary condition for the imaginary time.
The resultant quantity on the RHS of (\ref{integral}) is positive definite,
and therefore the numerical study by the MC simulation can be done
without any difficulty.
It should be remarked that the Lagrange multiplier $\lambda_i$
in Eq.(\ref{integral}) behaves as a gauge field, i.e.,
the RHS of (\ref{integral}) is invariant under the following ``gauge transformation",
$\omega_{\sigma,i}\rightarrow \omega_{\sigma,i}+\alpha_i, \ 
\lambda_i \rightarrow \lambda_i-\partial_\tau \alpha_i$.
In the practical calculation, we shall show that all physical quantities are
invariant under the above gauge transformation.

Various phases can form in the system (\ref{HtJ}) at $T=0$, e.g., states with 
long-range spin orders (FM, AF, spiral, etc), superfluid with Bose condensation
of $a$ and/or $b$ atoms, and superposition of them 
(the supersolid (SS))\cite{KKI,altman,sansone,phases}.
In the following sections, we shall consider several specific cases of the extended 
B-t-J model $H_{\rm tJ}+H_{\rm V}$ and derive effective field-theory model.
Then we clarify the phase diagram of the B-t-J model
by studying the effective model.

%%%%%%%%%%%%%%%%%%%%%%%%%%%%
\setcounter{equation}{0}
\section{Phase diagram of the extended quantum XY model: numerical study}
\subsection{FM coupling on square lattice}

In this section, we shall study the effective field theory obtained in the
previous section on the two-dimensional (2D) square lattice as the first example.
One of the simplest case is $xy$-FM system corresponding to
$J_z=0$ and $J<0$ in Eq.(\ref{HtJ}), though an extension to 
the case $|J_z|<-J$ is rather straightforward.
In this case the pseudo-spin symmetry is $O(2)$,  and then we put
$\rho_{1i}=\rho_{2i}=\rho$ and $\rho_{3i}=1-2\rho$.
The partition function $Z_{\rm qXY}$ of the effective theory is 
obtained as follows from Eqs.(\ref{Z}) and (\ref{integral}),
\begin{equation}
Z_{\rm qXY}=\int[\prod_{\sigma=1,2,3} D\omega_\sigma]
\exp\Big[-A_\tau-A(e^{i\Omega_\sigma},e^{-i\Omega_\sigma})\Big],
\label{Z2}
\end{equation}
where
\begin{eqnarray}
&&A_\tau={1 \over 4V_0}\int d\tau\sum_{i,\sigma} 
(\partial_\tau \omega_{\sigma i}+\lambda_i)^2, \nonumber \\
&&A(e^{i\Omega_\sigma},e^{-i\Omega_\sigma})=\int d\tau\Big[ \nonumber \\
&& -\sum_{\langle i,j\rangle}
c_{\rm h}\Big(\cos (\Omega_{2,i}-\Omega_{2,j})+
\cos (\Omega_{3,i}-\Omega_{3,j})\Big) \nonumber \\
&&\hspace{1cm}-c_{\rm s}\sum_{\langle i,j\rangle}\cos (\Omega_{1,i}-\Omega_{1,j})\Big]
\label{A1}
\end{eqnarray}
where 
$$
\Omega_{1i}=\omega_{1 i}-\omega_{2 i}, \
\Omega_{2i}=\omega_{1 i}-\omega_{3 i}, \
\Omega_{3i}=\omega_{2 i}-\omega_{3 i},
$$
and
\begin{equation}
c_{\rm h}={t \over 2}\rho(1-2\rho), \;\; c_{\rm s}=4J\rho^4.
\label{fg}
\end{equation}
It is obvious that the integrand in Eq.(\ref{Z2}) is positive definite,
and therefore the numerical calculation of $Z_{\rm qXY}$ by means of the
MC simulation is possible without any difficulty.
The model (\ref{Z2}) is a kind of lattice rotor model.

For practical calculation, we introduce a lattice for the imaginary-time 
direction and use the following lattice action $A_{L\tau}$ corresponding to $A_\tau$
\begin{eqnarray}
A_{{\rm L}\tau}=c_\tau\sum_{r} \sum_{\sigma=1}^3
\cos (\omega_{\sigma,r+\hat{\tau}}-\omega_{\sigma,r}+\lambda_r),  
\label{Atau}
\end{eqnarray}
where $r$ denotes site of the space-time cubic lattice,
$c_\tau={1 \over V_0\Delta \tau}$ and $\Delta \tau$ is the lattice
spacing of the imaginary-time direction.

We numerically studied the lattice model defined by the lattice action
\begin{equation}
A_{\rm Lxy}=A_{{\rm L}\tau}+A_{\rm L}(e^{i\Omega_\sigma},e^{-i\Omega_\sigma}), 
\label{AL1}
\end{equation}
where
$A_{\rm L}(e^{i\Omega_\sigma},e^{-i\Omega_\sigma})$ 
corresponds to $A(e^{i\Omega_\sigma},e^{-i\Omega_\sigma})$ in Eq.(\ref{A1}),
\begin{eqnarray}
&&A_{\rm L}(e^{i\Omega_\sigma},e^{-i\Omega_\sigma})= \nonumber \\
&& -\sum_{\langle r,r'\rangle}
C_3\Big(\cos (\Omega_{2,r}-\Omega_{2,r'})+
\cos (\Omega_{3,r}-\Omega_{3,r'})\Big) \nonumber \\
&&\hspace{1cm}-C_1\sum_{\langle r,r'\rangle}\cos (\Omega_{1,r}-\Omega_{1,r'}),
\label{AL2}
\end{eqnarray}
where $C_1=c_{\rm s}\Delta \tau$ and $C_3=c_{\rm h}\Delta \tau$, and 
$\langle r,r'\rangle$ denotes the NN sites in the 2D spatial lattice.
Here it is helpful to notice that
the parameters $C_1\propto {J/V_0}$ and 
$C_3\propto {t/V_0}$ are dimensionless.
We fixed the value of the dimensionless parameter
$c_\tau$ and calculated the partition function $Z$, 
the ``internal energy" $E$
and ``specific heat" $C$ as a function of $C_1$ and $C_3$,
\begin{eqnarray}
Z&=&\int [d\omega]e^{-A_{\rm Lxy}}, \nonumber \\
E&=&\langle A_{\rm Lxy} \rangle/L^3, \nonumber  \\
C&=&\langle (A_{\rm Lxy}-E)^2 \rangle/L^3,
\label{EC}
\end{eqnarray}
where $L$ is the linear size of the 3D cubic lattice.
In order to identify various phases, we also calculated the following
pseudo-spin and boson correlation functions,
\begin{eqnarray}
&& G_{\rm S}(r)={1 \over L^3}\sum_i \langle 
e^{i\Omega_{1,i}}e^{-i\Omega_{1,i+r}}\rangle, \nonumber \\
&& G_{a}(r)={1 \over L^3}\sum_i \langle 
e^{i\Omega_{2,i}}e^{-i\Omega_{2,i+r}}\rangle, \nonumber \\
&&G_{b}(r)={1 \over L^3}\sum_i \langle 
e^{i\Omega_{3,i}}e^{-i\Omega_{3,i+r}}\rangle,
\label{CF1}
\end{eqnarray}
where sites $i$ and $i+r$ are located in the same spatial 2D lattice,
i.e., the equal-time correlations.
For example, $G_a(r)\rightarrow\mbox{finite}$ as $r\rightarrow \infty$
indicates Bose-Einstein condensation (BEC) of the $a$-atom.

For numerical simulations, 
we employ the standard Monte-Carlo Metropolis algorithm
with local update\cite{Met}.
The typical sweeps for measurement is $(30000 \sim 50000)\times (10$ samples),
and the acceptance ratio is $40\% \sim 50\%$.
Errors are estimated from 10 samples with the jackknife methods.

%%%%%%%%%%%%%%%%%%%%%%%%%%%%%%%%%%%%%%%%%%%%%%%%%%%%%%%%%%%
%FIG.1
\begin{figure}[t]
\begin{center}
\includegraphics[width=4cm]{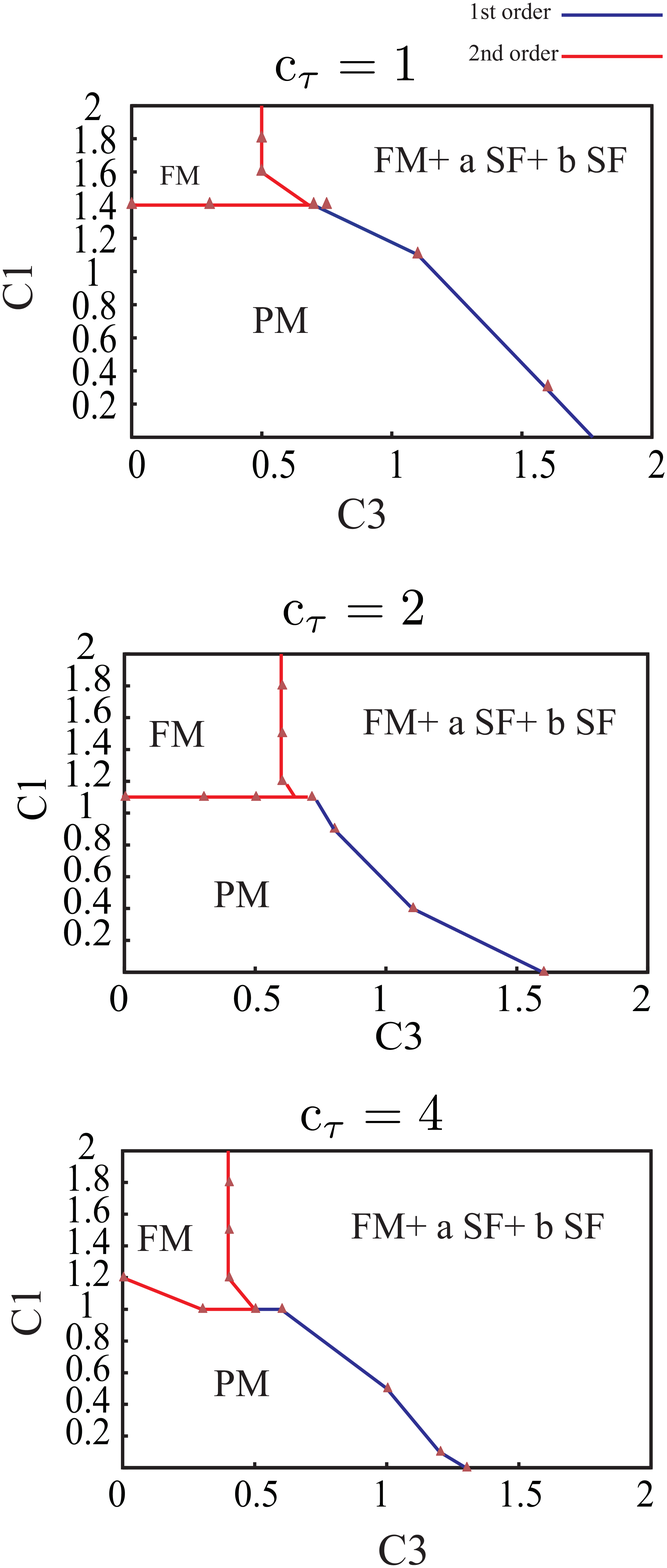}
\includegraphics[width=4.2cm]{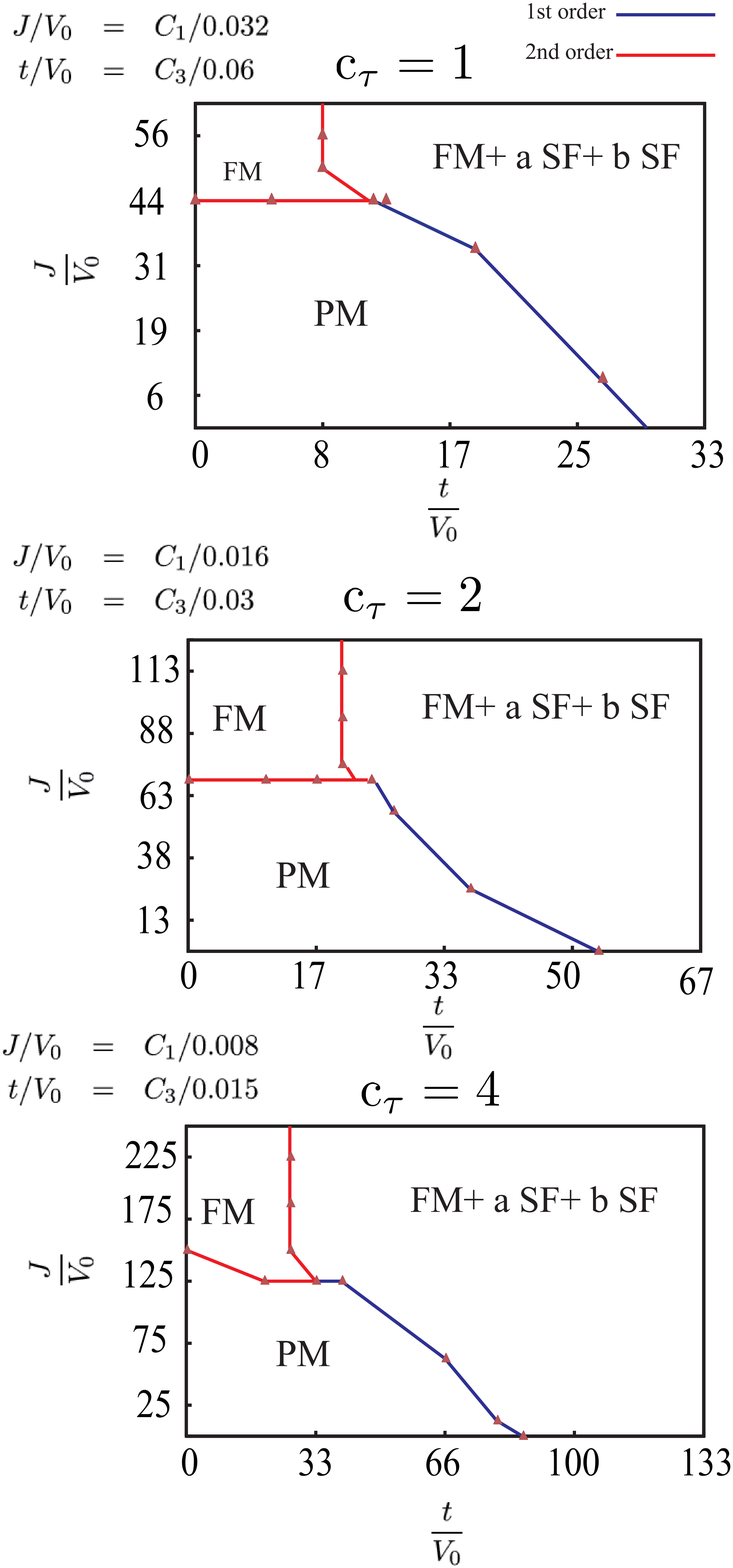}
\vspace{-0.3cm}
\caption{(Color online)
Phase diagram of the effective field theory $A_{\rm Lxy}$
on 3D cubic lattice for various values of $c_\tau$. 
$\rho=0.35$.
Dots denote the observed phase transition point.
Thick line denotes first-order phase transition line whereas the others are
second-order transition lines.
System size $L=24$.
}\vspace{-0.5cm}
\label{PD1}
\end{center}
\end{figure}
%%%%%%%%%%%%%%%%%%%%%%%%%%%%%%%%%%%%%%%%%%%%%%%%%%%%%%%%%%%
%%%%%%%%%%%%%%%%%%%%%%%%%%%%%%%%%%%%%%%%%%%%%%%%%%%%%%%%%%%
%FIG.2
\begin{figure}[t]
\begin{center}
\includegraphics[width=7cm]{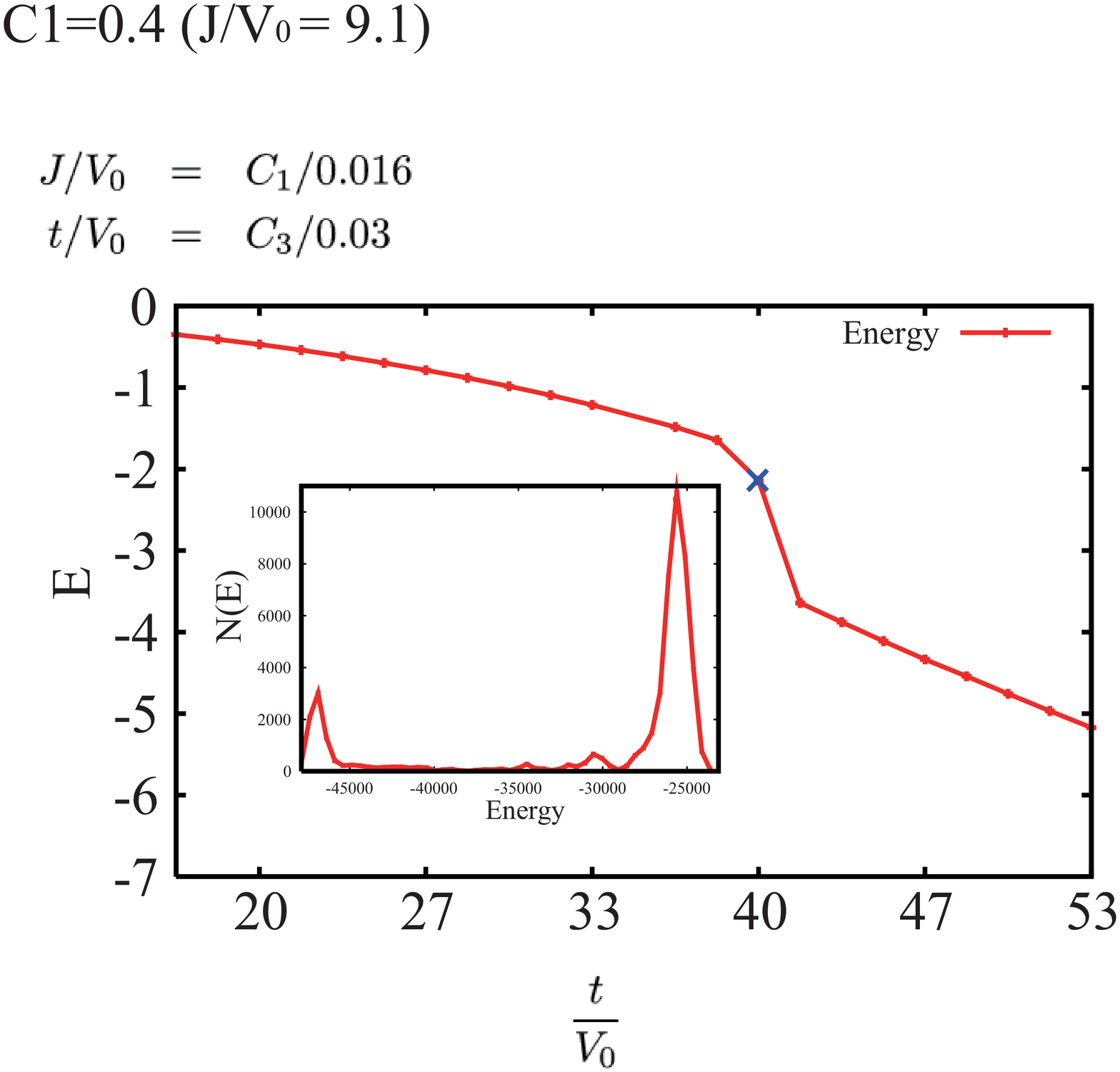}
\vspace{-0.3cm}
\caption{(Color online)
Internal energy $E$ and density of state $N(E)$ for the phase transitions shown in 
Fig.\ref{PD1}.
Double-peak shape for ${J\over V_0}=9.1$ and ${t\over V_0}=40.0$ indicates 
first-oder phase transition.
$c_\tau=2$ and $L=24$.
}
\vspace{-0.5cm}
\label{fig:NE}
\end{center}
\end{figure}
%%%%%%%%%%%%%%%%%%%%%%%%%%%%%%%%%%%%%%%%%%%%%%%%%%%%%%%%%%%
%%%%%%%%%%%%%%%%%%%%%%%%%%%%%%%%%%%%%%%%%%%%%%%%%%%%%%%%%%%
%FIG.3
\begin{figure}[t]
\begin{center}
\vspace{0.3cm} \includegraphics[width=5cm]{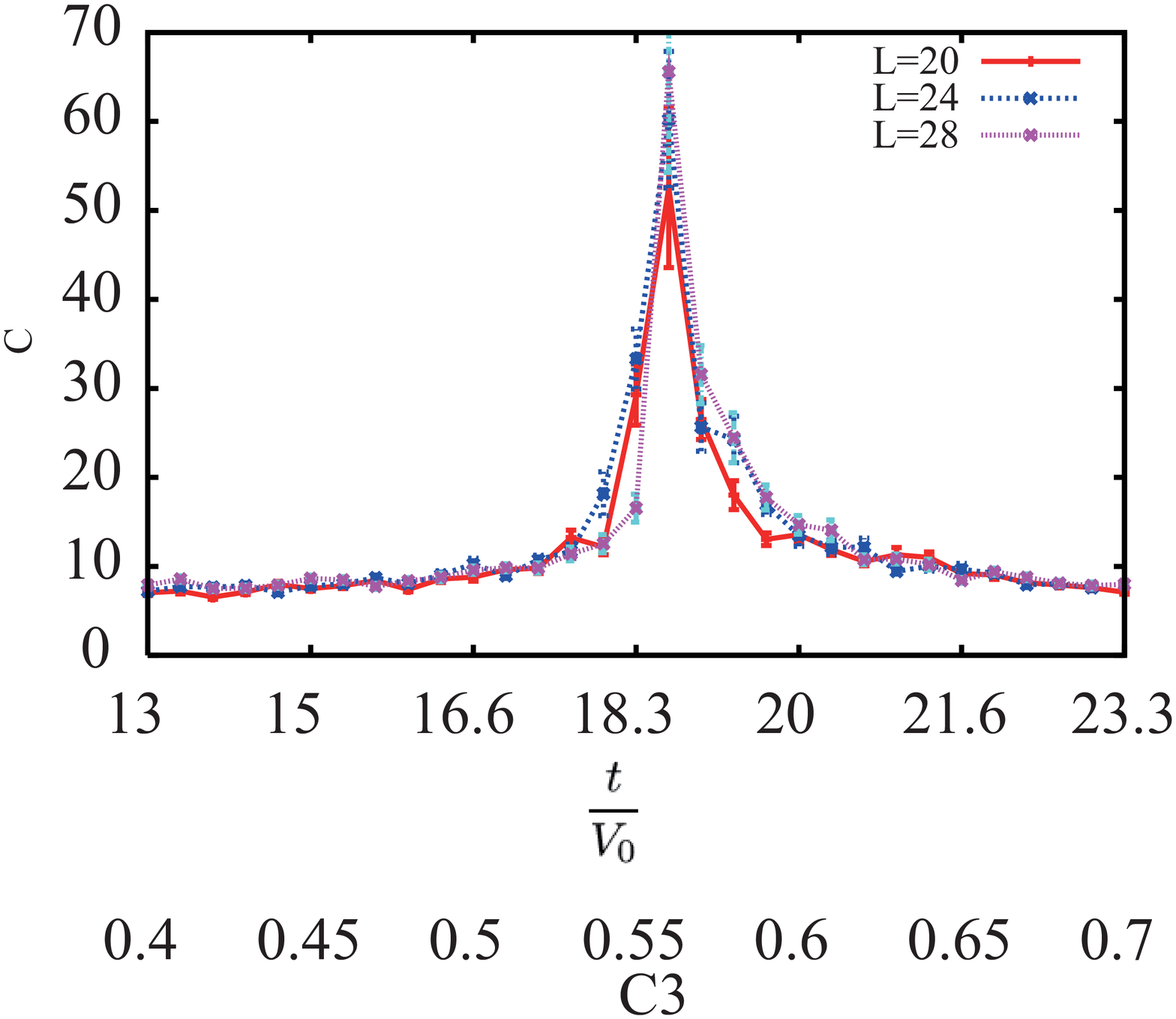} 
\vspace{0.5cm} \\
\includegraphics[width=5cm]{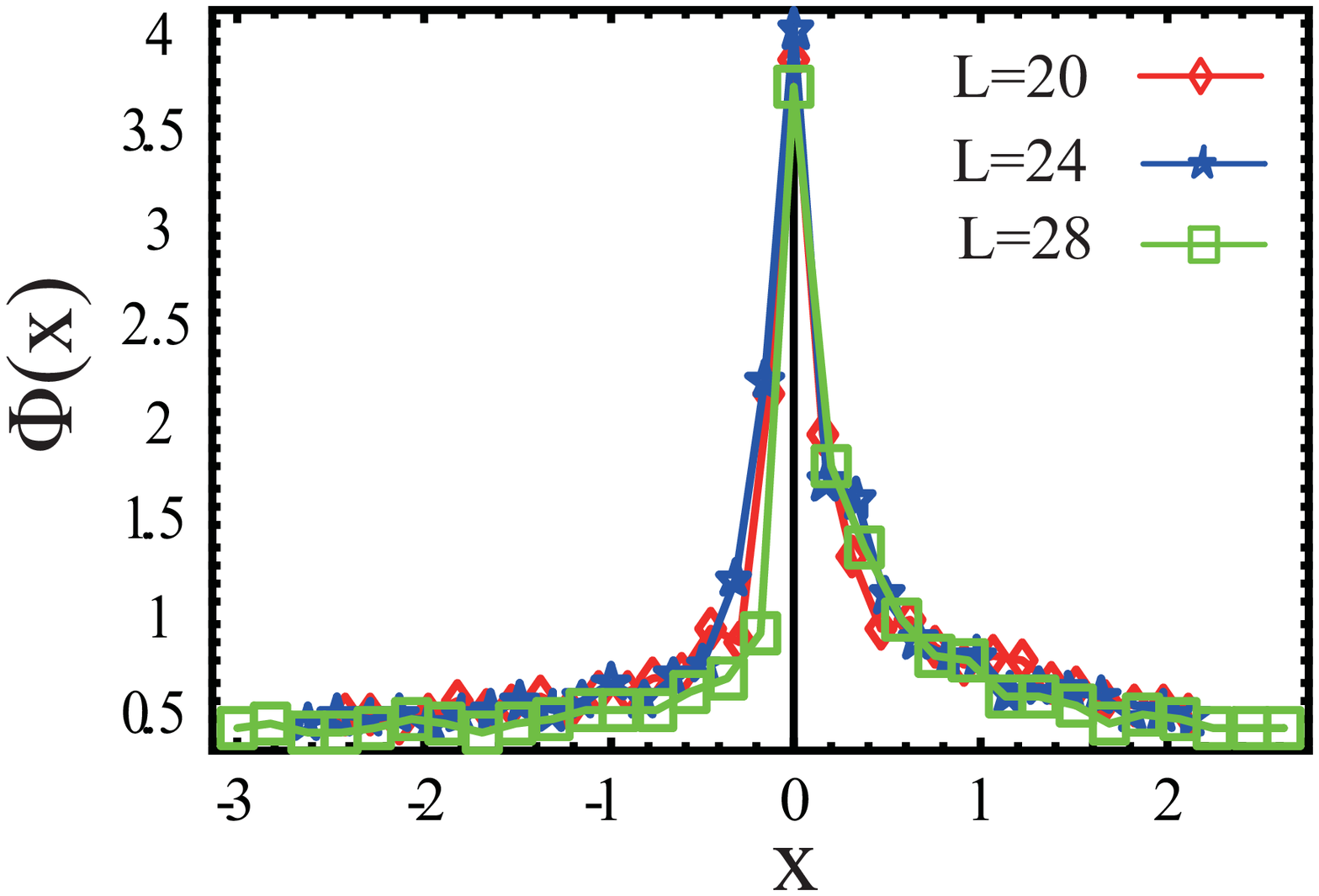} 
\vspace{-0.7cm}
\caption{(Color online)
Specific heat $C$ and its finite-size scaling function $\Phi(x)$ 
for the second-order phase transition shown in Fig.\ref{PD1}.
$c_\tau=2$ and ${J\over V_0}=87.5(C_1=1.4)$.
}
\vspace{-0.5cm}
\label{fig:FSS}
\end{center}
\end{figure}
%%%%%%%%%%%%%%%%%%%%%%%%%%%%%%%%%%%%%%%%%%%%%%%%%%%%%%%%%%%

We numerically studied the model with the system size $L=14, 16, 20, 24, 28$ and
show the obtained phase diagram for $\rho=0.35$ in Fig.\ref{PD1} using
$L=24$ data.
There are three phases and they are separated with each other by 
first or second-order phase transition lines.
To identify the first-order phase transition, we calculated the ``density of state"
$N(E)$ defined by
\begin{eqnarray}
N(E)\Delta E&=&\mbox{Number of configurations with} \nonumber \\
&& A_{\rm Lxy}=(E+\Delta E)\sim E.
\label{NE}
\end{eqnarray}
Step function like behavior of $E$ and
double-peak shape of $N(E)$ at the critical point indicate the
existence of a first-order phase transition.
We show some of the results in Fig.\ref{fig:NE}.
On the other hand to identify second-order phase transition,
we used nonsingular behavior of $E$, single-peak shape of $N(E)$ and 
system-size dependence of the specific heat $C$ in Eq.(\ref{EC}).
For example, see Fig.\ref{fig:FSS}.
System-size dependence of the specific heat is parameterized as follows,
\begin{equation}
C_L(\epsilon)=L^{\sigma/\nu}\Phi(L^{1/\nu}\epsilon),
\label{FSS}
\end{equation}
where $\nu$ and $\sigma$ are critical exponents, 
$\epsilon=(C_3-C_{3\infty})/C_{3\infty}$ with $C_{3\infty}=$ the critical coupling of 
$L\rightarrow \infty$, and $\Phi(x)$ is the scaling function.
Estimated values are $\nu=1.35, \ \sigma=1.2$ and 
$C_{3\infty}=0.56 \ ({t_\infty\over V_0}=19)$.

From Fig.\ref{PD1},
it is obvious that for small $t$ and $J$, there exists a ``paramagnetic state"
(PM state) without any long-range orders
(LRO's) and its domain in the $c_3-c_1$ plane is decreased for increasing $c_\tau$.
In this PM state, a superposition of $a$-atom and $b$-atom is realized
at each site, but coherence of the relative phase in the superposition does
not exists.
It should be notice that the atomic BEC always accompanies the pseudo-spin LRO.
The obtained results are in good agreement with the result of the previous study on
some related model on 3D cubic lattice at finite $T$\cite{BtJ1}.

%%%%%%%%%%%%%%%%%%%%%%%%%%%%%%%%%%%%%%%%%%%%%
\subsection{AF coupling on triangular lattice}

In this subsection, we shall study the B-t-J model on the 2D triangular lattice.
As we also discretize the imaginary time, the model $A_{\rm Lxy}$ in Eq.(\ref{AL1}) is
defined on the 3D stacked triangular lattice.
In this section we consider the case of the $xy$-AF case with $O(2)$ symmetry.
Therefore we set $J>0$ and $J_z=0$ in Eq.(\ref{HtJ}).
As there exists the frustration, we expect that various phases appear
in contrast to the $xy$-FM case.
In later section, we also show the result of the numerical study of the B-t-J model
on the stacked triangular lattice {\em at finite but low T}, 
which is closely related to the 2D model  at $T=0$.
This close relation between 3D model at low $T$ and 2D model at $T=0$
has been previously observed in various systems\cite{models}.
Origin of this similarity is somewhat obvious from $A_{{\rm L}\tau}$ 
in Eq.(\ref{Atau}) as it can be regarded as an inter-layer coupling.

To perform the numerical calculation, we first assign values of the parameters
$\rho_{1i}, \rho_{2i} \ (\rho_{3i}=1-\rho_{1i}-\rho_{2i})$.  
Here we assume a homogeneous state and put $\rho_{1i}=\rho_{2i}=\rho
\ (\rho_{3i}=1-2\rho)$ as in the previous case.
Results of more general cases will be reported in a future publication.

\begin{figure}[t]
\begin{center}
\includegraphics[width=5.5cm]{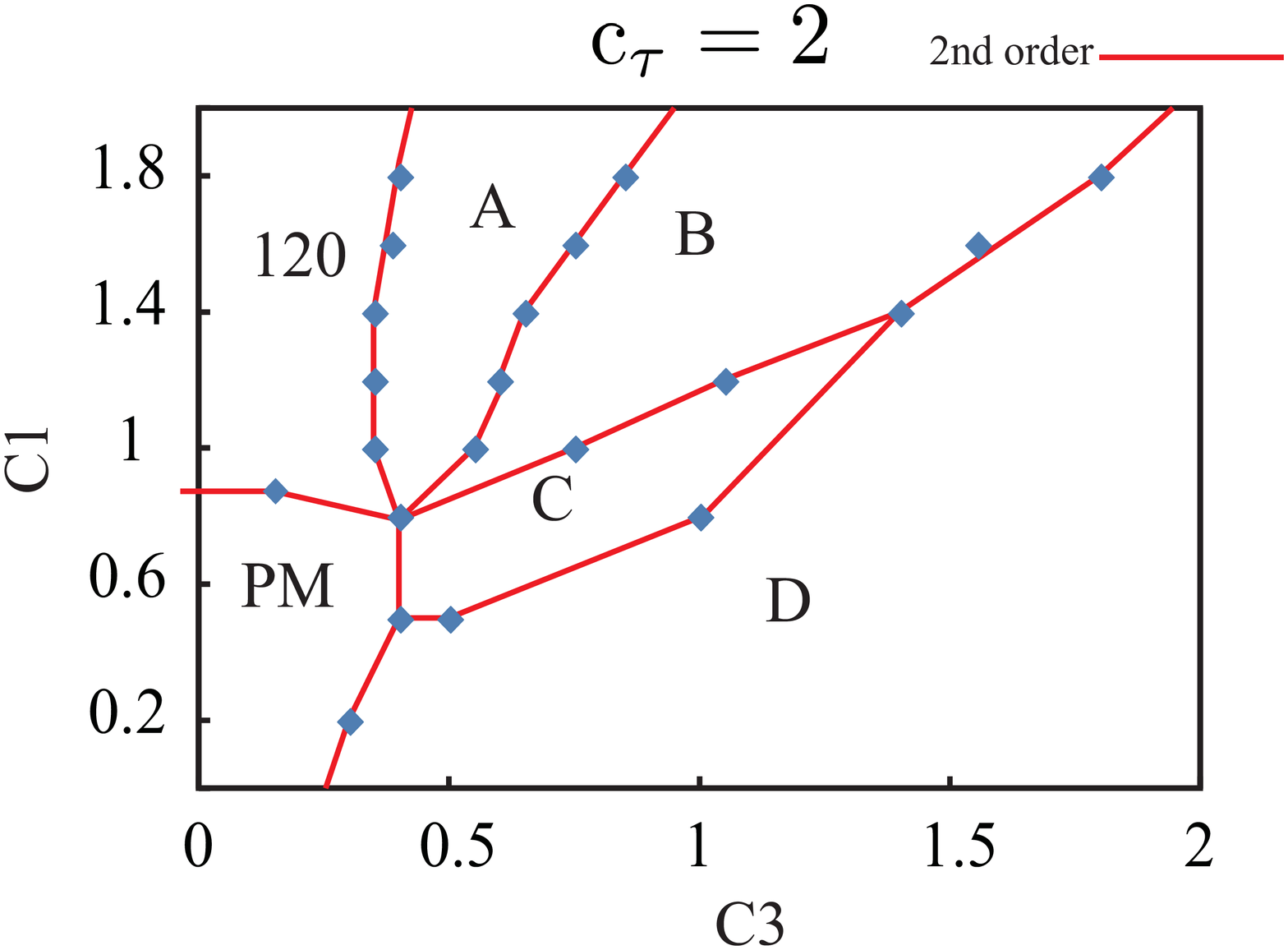}
\includegraphics[width=5.5cm]{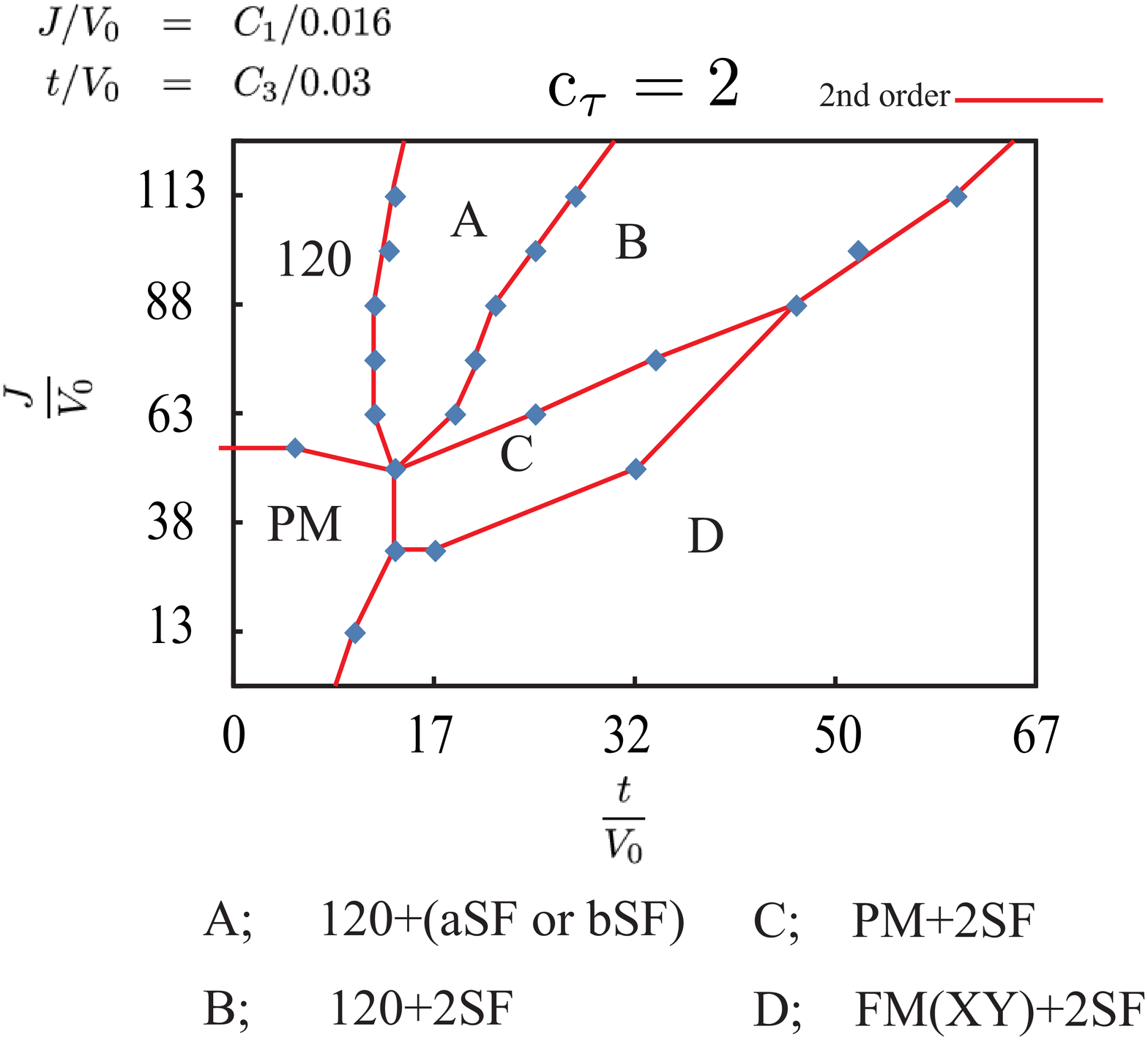}
\vspace{-0.3cm}
\caption{(Color online)
Phase diagram of the quantum XY model on triangular lattice. 
Density of each boson $\rho=0.3$.
System size $L=18$.
There are six phases. Phase of $120^o$ denotes the state with $120^o$
pseudo-spin long-range order without atomic BEC.
SF stands for superfluid.
}\vspace{-0.5cm}
\label{PD2}
\end{center}
\end{figure}

\begin{figure}[h]
\begin{center}
\includegraphics[width=4.5cm]{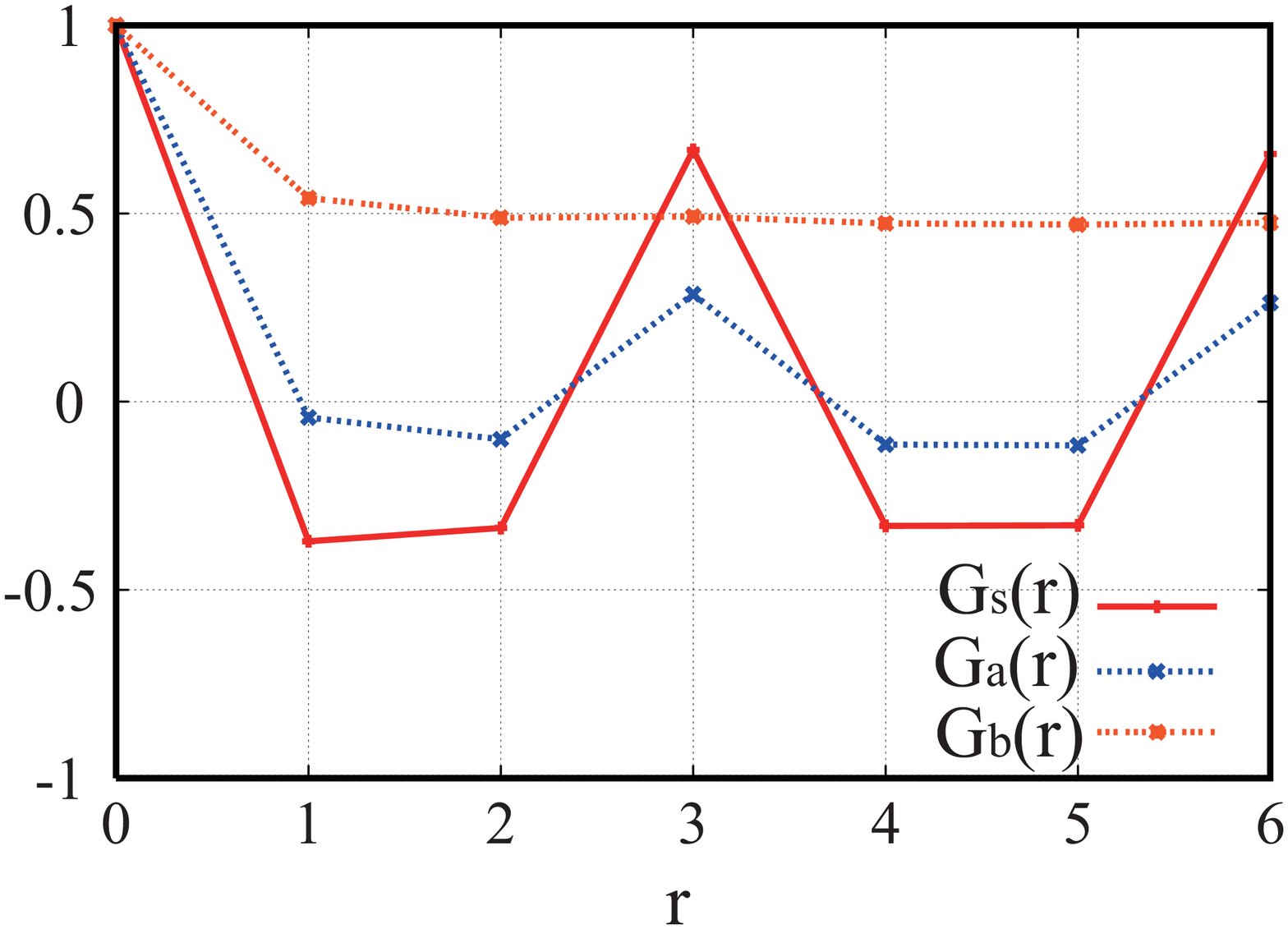}
\includegraphics[width=4cm]{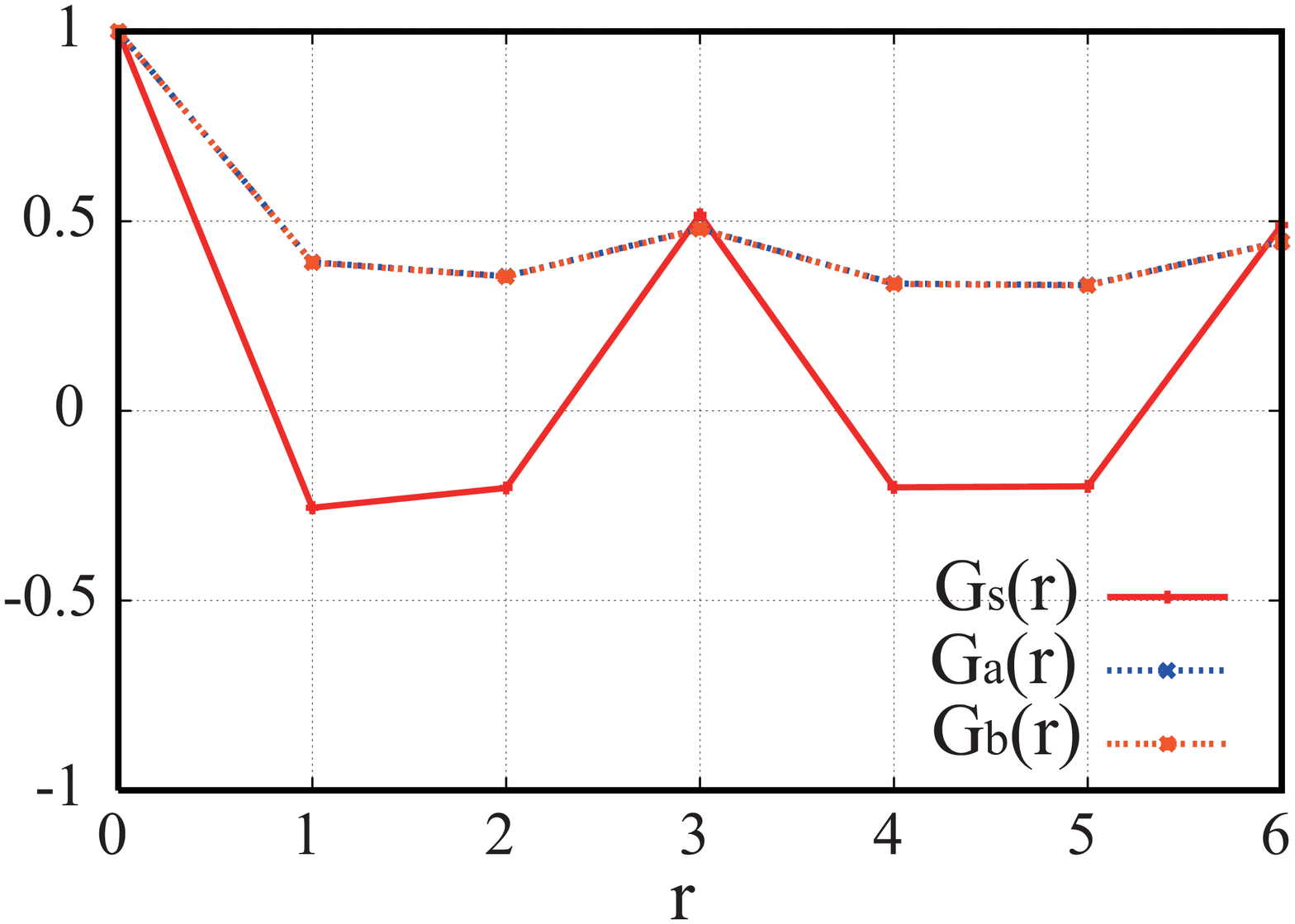}
\vspace{-0.3cm}
\caption{(Color online)
Correlation functions in the phase diagram.
$t/V_0=21.7(C_3=0.65)$ and $J/V_0=100(C_1=1.6)$ (left), 
$t/V_0=33.3(C_3=1.0)$ and $J/V_0=100(C_1=1.6)$ (right).
$L=12$.
}\vspace{-0.5cm}
\label{corF1}
\end{center}
\end{figure}

We numerically studied the system on the triangular lattice for
the system size $L=12, 14$ and $18$.
In Fig.\ref{PD2}, we show the obtained phase diagram for $\rho=0.3$
and the hole density$=0.4$.
Order of the phase transitions were determined as in the case of the
FM coupling on the square lattice.
There are six phases in the $t-J$ plane, and
the order of the phase transitions was determined as in the case of the
square lattice.
For small $t$ (i.e., small $C_3$), 
the pseudo-spin degrees of freedom exhibits the 120$^o$
long-range order.
As we assume a homogeneous hole distribution, a superposed state of 
atoms and hole is realized at each site, but coherent condensation of
atoms does not takes place yet.
As $t$ is increased, phase transition to the states with the 120$^o$
long-range order and Bose condensation take place.
For example in the phase B in Fig.\ref{PD2}, the both atoms Bose
condense, as the correlation functions shown in Fig.\ref{corF1} exhibit. 
In the mean-field approximation, the wave function of that state is given
by
\begin{eqnarray}
|\Psi\rangle&&=\prod_{i\in A}[a^\dagger_i+b^\dagger_i+c_A] \nonumber \\ \nonumber
&&\times\prod_{i\in B}[a^\dagger_i+e^{i{2\pi \over 3}}b^\dagger_i+c_B]  \\ \nonumber
&&\times\prod_{i\in C}[a^\dagger_i+e^{-i{2\pi \over 3}}b^\dagger_i+c_C]|0\rangle,
\end{eqnarray}
where $c$'s are some complex number and $|0\rangle$ is the empty state
of $a$ and $b$-atoms.
This phase will be discussed by using the effective field theory in Sec.IV.C
and also by the MC simulations in Sec.V.
As the parameter $C_3$ is increased further, order of the pseudo-spin
is destroyed first (phase C) and then changes to the FM one in the
 $S^x-S^y$ plane (phase D).

%%%%%%%%%%%%%%%%%%%%%%%%%%%%%%%%%%%%%%%%%%%%%%%%%%%%%%%%%%%
%FIG.4
%%%%%%%%%%%%%%%%%%%%%%%%%%%%%%%%%%%%%%%%%%%%%%%%%%%%%%%%%%%
%%%%%%%%%%%%%%%%%%%%%%%%%%%%%%%%%%%%%%%%%%%%%%%%%%%%%%%%%%%
%FIG.5
%%%%%%%%%%%%%%%%%%%%%%%%%%%%%%%%%%%%%%%%%%%%%%%%%%%%%%%%%%%

%%%%%%%%%%%%%%%%%%%%%%%%%%%%
\setcounter{equation}{0}
\section{Effective field theory for Bose condensation and pseudo-spin order}

In the previous sections, we numerically studied lattice quantum XY model
that describe the two-component bosons with strong repulsions.
The obtained phase diagrams show that there exist Bose-condensed
phases as well as the state of simple pseudo-spin LRO.
In this section, we shall derive an effective field theory from the
quantum XY model, which describes directly the Bose condensation and spin order.
This field theory not only explains the numerically obtained phase diagram but
also reveals low-energy excitations and interactions between them.
This kind of effective field theory has been obtained and discussed for a granular 
superfluid etc by using Hubbard-Stratonovichi transformation\cite{HS}.
In this section we shall employ a slightly different method\cite{source}.
As a result, we reveal some important point that
has been overlooked so far.

%%%%%%%%%%%%%%%%%%%%%%%%%%%
\subsection{Effective field theory}

To illustrate the procedure to derive the effective field theory,
we first consider a single rotor model as an example that describes
superfluid phase transition.
After consideration of the simple model, we shall apply similar methods
to the quantum XY model for the B-t-J model.

We first rewrite the partition function of the rotor model on the square
lattice by introducing
source terms,
\begin{eqnarray}
Z_{\rm rotor}&=&\int [d\omega]e^{-{1 \over V_o}
\int d\tau \sum \dot{\omega}^2_{i}
-\int d\tau H(e^{i\omega_{i}}, e^{-i\omega_{i}}) }
\nonumber \\
&=&\int [d\omega]e^{-{1 \over V_o}
\int d\tau \sum \dot{\omega}^2_{i}
-\int d\tau 
H({\delta \over \delta\eta_{i}}, {\delta \over \delta\bar{\eta}_{i}})}
\nonumber  \\
&& \times
I(\eta_{i}, \omega_{i})|_{\eta=\bar{\eta}=0},  \label{rotorZ} \\
I(\eta_{i}, \omega_{i})&=&
e^{\int d\tau\sum (\eta_{i}e^{i\omega_{i}}+
\bar{\eta}_{i}e^{-i\omega_{i}})},
\label{sourcerotor}
\end{eqnarray}
where $\dot{\omega}_{i}=\partial_\tau {\omega}_{i}$ and 
$$H(e^{i\omega_{i}}, e^{-i\omega_{i}})=-J\sum_{\langle i, j \rangle}
\cos(\omega_i-\omega_j).
$$ 
In Eq.(\ref{rotorZ}), we evaluate the path integral of $\omega_i$ as
\begin{eqnarray}
&& \int [d\omega]e^{-{1 \over V_0}
\int d\tau \sum_\sigma \dot{\omega}^2_{i}}I(\eta_{i}, \omega_{i}) \nonumber \\
&& \hspace{1cm} =e^{\int d\tau \int d\tau'\bar{\eta}_{i}(\tau)
e^{-V_0|\tau-\tau'|}\eta_{i}(\tau')},
\label{intomega}
\end{eqnarray}
where we have used 
\begin{eqnarray}
\langle e^{i\omega_{i}(\tau)} e^{-i\omega_{i}(\tau')} \rangle 
=e^{-V_0|\tau-\tau'|},
\label{G}
\end{eqnarray}
and the fact that other correlators like 
$\langle e^{i\omega_{i}(\tau)} e^{i\omega_{i}(\tau')} \rangle $
are vanishing.
RHS of Eq.(\ref{intomega}) can be expressed by introducing a complex
boson field $\Phi_i$ as 
\begin{eqnarray}
&& e^{\int d\tau \int d\tau'\bar{\eta}_{i}(\tau)
e^{-V_0|\tau-\tau'|}\eta_{i}(\tau')} \nonumber \\
&& =\int[d\Phi]\exp \Big[-{1 \over V_0}\int d\tau
\Phi^\ast_{i}(-\partial^2_\tau+V^2_0)\Phi_{i} \nonumber \\
&& \hspace{0.5cm}
+\int d\tau(\eta_{i}\Phi_{i}+\bar{\eta}_{i}\Phi^\ast_{i})\Big].
\label{Phi}
\end{eqnarray}
By inserting Eq.(\ref{Phi}) into Eq.(\ref{rotorZ}), we obtain the
effective field theory of the rotor model with the following action $A_{\rm rotor}$,
\begin{eqnarray}
A_{\rm rotor}&=&\int d\tau \Big[{1 \over V_0}\sum_i
\Phi^\ast_{i}(-\partial^2_\tau+V^2_0)\Phi_{i} \nonumber \\
&& -{J\over 2}\sum_{\langle i, j\rangle}(\Phi^\ast_i\Phi_j+\mbox{c.c.})\Big], \nonumber \\
Z_{\rm rotor}&=&\int [d\Phi]e^{-A_{\rm rotor}}.
\label{A_Phi}
\end{eqnarray}

The above derivation from Eq.(\ref{rotorZ}) to 
Eq.(\ref{A_Phi}) seems exact because the
integration of $\omega_{i}(\tau)$ is the one-site integral and essentially
Gaussian integration of the free fields.
In fact the above manipulation is exact as long as the boson field 
$\Phi_{i}$ does not Bose condense.
On the other hand for $J \gg V_0$, the Bose condensation of $\Phi_{i}$
takes place, i.e., $\langle \Phi_{i} \rangle \neq 0$.
One may wonder if the above manipulation is applicable even
for this case because Eq.(\ref{G}) seems to indicate {\em nonexistence} 
of the long-range oder of  $\Phi_{i}$.
Furthermore in this case, the mass term of $\Phi_{i}$ becomes negative,
and the integration of $\Phi_{i}$ in Eq.(\ref{A_Phi}) becomes unstable, e.g.,
for the square lattice
\begin{eqnarray}
&& {J \over 2}\sum_{i,\mu}(\Phi^{\ast}_{i}\Phi_{i+\mu}+
\mbox{c.c.})-V_0\sum_i|\Phi_i|^2
\nonumber \\
&& \;\;\;=-{J \over 2}\sum_{i,\mu}|\nabla _{\mu}\Phi_{i}|^2
+\sum_{i}(zJ-V_0)|\Phi_{i}|^2,
\end{eqnarray}
where $\nabla _{\mu}$ is the lattice difference operator and $z$ is the number
of links emanating from a single site and $z=4$ for the square lattice.
This instability comes from the fact that the order of the $\eta$-derivative
and the $\omega$-integration is {\em not interchangeable} when
the Bose condensation, i.e., a phase transition to an ordered state, takes place.
It is obvious 
$\langle \Phi_{i} \rangle=\langle e^{i\omega_{i}} \rangle <1$
in the Bose condensed state,
and therefore it is plausible to expect term like $\lambda|\Phi_{i}|^4$
to appear in the effective field theory to stabilize the integration of 
$\Phi_{i}$ for the Bose condensed state, though explicit calculation
to determine the coefficient is difficult.
Simple spin-wave like approximation for the rotor model,
$$
J\cos(\omega_i-\omega_j)\sim J-{J \over 2}(\omega_i-\omega_j)^2,
$$
gives the estimation like 
$\langle e^{i\omega_{i}} \rangle\langle e^{-i\omega_{i}} \rangle
\sim e^{-{V_0\over J}}$, and then
the coefficient of the $|\Phi_{i}|^4$-term is estimated as 
$\lambda \sim J e^{{V_0\over J}}$ for $J\gg V_0$.

Hubbard-Stratonovich transformation derives a similar effective field theory
to the above.
But its straightforward application gives a negative coefficient of the 
$|\Phi_i|^4$-term indicating an instability of the system.
This means that certain step of the derivation is invalid, e.g.,
introduction of the Hubbard-Stratonovichi field and the $\omega$-integration
is not interchangeable.
This problem is under study and the result will be reported in a separate
publication.

Similar manipulation to the above can be applied to the present quantum extended
XY model.
From Eqs.(\ref{Z2}) and (\ref{A1}), 
\begin{eqnarray}
Z_{\rm qXY}&=&\int [d\omega]e^{-{1 \over V_o}
\int d\tau \dot{\omega}^2_{\sigma i}
-A(e^{i\Omega_{\sigma i}}, e^{-i\Omega_{\sigma i}}) }
\nonumber \\
&=&\int [d\omega]e^{-{1 \over V_o}
\int d\tau \sum \dot{\omega}^2_{\sigma i}
-A({\delta \over \delta\eta_{\sigma i}}, {\delta \over \delta\bar{\eta}_{\sigma i}})}
\nonumber  \\
&& \times
I(\eta_{\sigma i}, \Omega_{\sigma i})|_{\eta=\bar{\eta}=0}, \nonumber \\
I(\eta_{\sigma i}, \Omega_{\sigma i})&=&
e^{\int d\tau\sum (\eta_{\sigma i}e^{i\Omega_{\sigma i}}+
\bar{\eta}_{\sigma i}e^{-i\Omega_{\sigma i}})},
\label{source}
\end{eqnarray}
where $\dot{\omega}_{\sigma i}=\partial_\tau {\omega}_{\sigma i}$ and 
we have omitted the gauge field $\lambda_i$ as we consider the
only gauge-invariant objects through $\Omega_{\sigma i}$.
Then the integration over $\omega_{\sigma i}$ can be performed as,
\begin{equation}
\int [d\omega]e^{-{1 \over V_0}
\int d\tau \sum_\sigma \dot{\omega}^2_{\sigma i}}I(\eta_{\sigma i}, \Omega_{\sigma i})
=\tilde{I}(\eta_{\sigma i}, \bar{\eta}_{\sigma i}),
\label{Lsource}
\end{equation}
and the Green function of $\omega_{\sigma i}(\tau)$, $G_\omega$, is obtained as
\begin{eqnarray}
\langle e^{i\omega_{\sigma i}(\tau)} e^{-i\omega_{\sigma' i}(\tau')} \rangle 
=\delta_{\sigma\sigma'}e^{-V_0|\tau-\tau'|}.
\label{G2}
\end{eqnarray}
Typical term of $\tilde{I}(\eta_{\sigma i}, \bar{\eta}_{\sigma i})$
is as follows,
$$
e^{\int d\tau \int d\tau'\bar{\eta}_{\sigma i}(\tau)
e^{-2V_0|\tau-\tau'|}\eta_{\sigma i}(\tau')},
$$
and this quantity is expressed
by introducing complex scalar fields $\Phi_{\sigma i}(\tau)$ as
\begin{eqnarray}
&& e^{\int d\tau \int d\tau'\bar{\eta}_{\sigma i}(\tau)
e^{-2V_0|\tau-\tau'|}\eta_{\sigma i}(\tau')} \nonumber \\
&& =\int[d\Phi]\exp \Big[-{1 \over 4V_0}\int d\tau
\Phi^\ast_{\sigma i}(-\partial^2_\tau+4V^2_0)\Phi_{\sigma i} \nonumber \\
&& \hspace{0.5cm}
+\int d\tau(\eta_{\sigma i}\Phi_{\sigma i}+\bar{\eta}_{\sigma i}\Phi^\ast_{\sigma i})\Big].
\label{Phisigma}
\end{eqnarray}
By inserting $\tilde{I}(\eta_{\sigma i}, \bar{\eta}_{\sigma i})$, which is expressed
in terms of the boson fields $\Phi_{\sigma i}(\tau)$ for $e^{i\Omega_{\sigma i}}$,
into Eq.(\ref{source}), the action of the effective field theory is obtained 
as follows, 
\begin{eqnarray}
A_0&=& \int d\tau\Big[\sum_{\sigma, \langle i,j\rangle}
(a_\sigma\Phi^\ast_{\sigma i} \Phi_{\sigma j})
-{1 \over 4V_0}\sum_{\sigma i}(|\dot{\Phi}_{\sigma i}|^2
+4V_0^2 |\Phi_{\sigma i}|^2) \nonumber   \\
&&+\sum_i g (\Phi^\ast_{1 i}\Phi_{2 i}\Phi^\ast_{3 i} +\mbox{c.c.})\Big],
\label{A0}
\end{eqnarray}
where $a_1=a_{\rm s}$, $a_2=a_3=a_{\rm h}$ and $g={3 \over 2V^2_0}$.
To describe the Bose condensed state,
we add the $|\Phi|^4$-terms in the effective field theory and discuss
the phase diagram of the system. 
The final form of the effective action is therefore given by
\begin{eqnarray}
A_{\rm eff}&=&
A_0-\int d\tau \sum_{\sigma, i}(\lambda_\sigma |\Phi_{\sigma i}|^4) \nonumber \\
&=&\int d\tau L_{\rm eff}, \nonumber \\
Z&=&\int [d\Phi]e^{-A_{\rm eff}}.
\label{Aeff}
\end{eqnarray}

%%%%%%%%%%%%%%%%%%%%%%%%%%%%
\subsection{Phase diagram and low-energy excitations: Square lattice}

In this subsection, we shall apply the field-theoretical approach
explained in the previous subsection to the B-t-J model on the square lattice
as a simple example.
Furthermore we focus on the FM parameter region $J<0$.
In this case there exists no frustration and therefore it is rather
straightforward to obtain the phase diagram.
Nevertheless study on this system reveals important aspect of 
the state with multiple long-range orders and structure of the
Nambu-Goldstone bosons.

%%%%%%%%%%%%%%%%%%%%%%%%%%%%%%%%%%%%%%%%%%%%%%%%%%%%%%%%%%%
%FIG.6
\begin{figure}[t]
\begin{center}
\includegraphics[width=7.2cm]{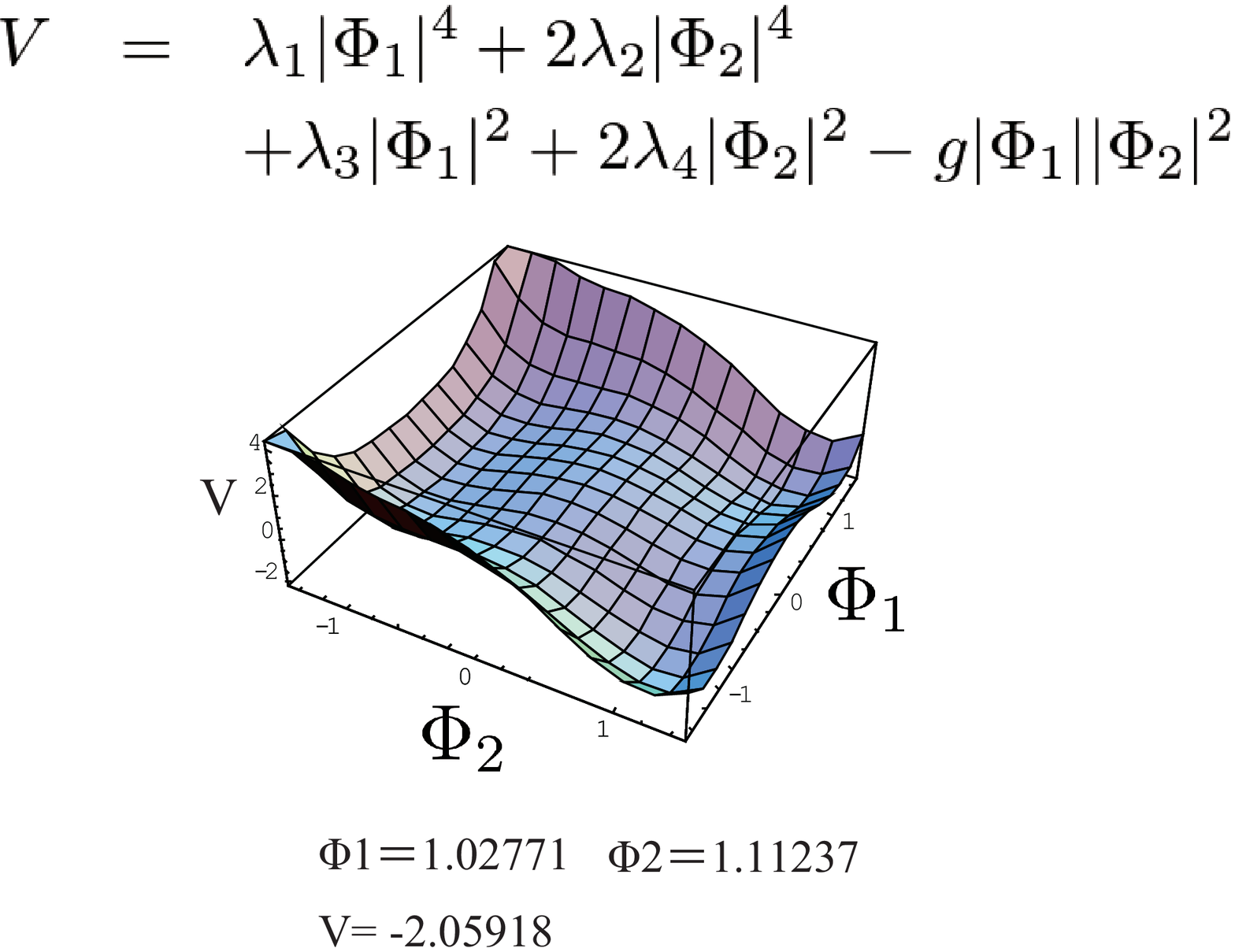}
\vspace{-0.3cm}
\caption{(Color online)
Potential $V(\Phi_\sigma)$ (\ref{V1}) and its minimum.
Symmetric case $\Phi_2=\Phi_3$.
Parameters are
$
\lambda_{1}=1,
\lambda_{2}=\frac{1}{2},
\lambda_{3}=-1,
\lambda_{4}=-1,$ and $g=1$.
}\vspace{-0.5cm}
\label{Pt1}
\end{center}
\end{figure}
%%%%%%%%%%%%%%%%%%%%%%%%%%%%%%%%%%%%%%%%%%%%%%%%%%%%%%%%%%%

This system was studied in Sec.III.A, and we obtained the phase diagram
by MC simulations.
The potential $V(\Phi_\sigma)$ of the present system is given as follows from the effective field theory in Sec.IV.A,
\begin{eqnarray}
V(\Phi_\sigma)&=&\sum_\sigma(V_0-2za_\sigma)|\Phi_{\sigma}|^2
-g (\Phi^\ast_{1}\Phi_{2}\Phi^\ast_{3} +\mbox{c.c.}) \nonumber \\
&&+\sum_{\sigma}\lambda_\sigma |\Phi_{\sigma}|^4.
\label{V1}
\end{eqnarray}
From the first term of Eq.(\ref{V1}), it is obvious that spontaneous
symmetry breaking occurs for $a_\sigma>V_0$, but the second terms
that represent the interplay between the order parameters give nontrivial
contribution to the phase diagram.
In Fig.\ref{Pt1}, we show typical potential and its minimum obtained 
from Eq.(\ref{V1}), which derives qualitatively the same phase structure with 
that obtained in Sec.III.A.

It is interesting to see how gapless low-energy excitations, i.e.,
Nambu-Goldstone (NG) bosons, appear in the present system.
In the phase with the FM spin order $\langle \Phi_3 \rangle=v > 0$, we set
$\Phi_{3i}=v+\psi_i+i\chi_i$, and then it is obvious that the field $\chi_i$ 
describes a NG boson that appears as a result of 
the spontaneous symmetry breaking of the U(1) pseudo-spin symmetry. 

Interesting point is that how many NG bosons appear in the phase with the
FM+2SF.
If $g=0$ in Eq.(\ref{A0}), it is obvious that there exist three NG bosons.
However in the original B-t-J model, the symmetry is U(1)$\times$U(1)
for the global phase rotation of $a$ and $b$-boson operators. 
In order to study the low-energy excitations in the effective field theory,
we shall take the continuum description instead of the lattice one though
it is not essential.

For the FM B-t-J model on the square lattice, it is straightforward to derive
effective Lagrangian $L_{\rm eff}$ in the continuum spacetime 
from the effective field theory on the lattice.
For example in Eq.(\ref{A0}), we put
\begin{eqnarray}
&& \sum_{i,\mu}\Phi^{\ast}_{i}\Phi_{i+\mu}+\Phi_{i}\Phi^{\ast}_{i+\mu}
\nonumber \\
&& \;\;\;=-\sum_{i,\mu}|\nabla _{\mu}\Phi_{i}|^2+\sum_{i}2z|\Phi_{i}|^2
\nonumber \\
&& \;\;\; \Rightarrow -\sum_{i,\mu} |\partial_\mu \Phi_i|^2
+\sum_i 2z|\Phi_{i}|^2.
\nonumber
\end{eqnarray}
Then $L_{\rm eff}$ in the continuum is derived as follows from Eq.(\ref{Aeff})
\begin{eqnarray}
L^{\rm sq}_{\rm eff}&=&\biggl(a_{s}
|\nabla \Phi_{1}(x)|^2+a_{h}|\nabla\Phi_{2}(x)|^{2}
+a_{h}|\nabla \Phi_{3}(x)|^{2}\biggr)\nonumber\\
&-&\biggl(\frac{1}{4V_{0}}|\partial_{\tau}\Phi_{1}(x)|^{2}+\frac{1}{4V_{0}}|\partial_{\tau}\Phi_{2}(x)|^{2}
 \nonumber \\
&&+\frac{1}{4V_{0}}|\partial_{\tau}\Phi_{3}(x)|^{2}\biggr)+V(\Phi_\sigma),
\label{Leff1}
\end{eqnarray}
where the potential $V(\Phi_\sigma)$ is given as
\begin{eqnarray}
V(\Phi_{\sigma})&=&(V_0-2za_{s})|\Phi_{1}(x)|^2 
+(V_0-2za_{h})|\Phi_{2}(x)|^2 \nonumber \\
&&+(V_0-2za_{h})|\Phi_{3}(x)|^2  \nonumber \\
&&-g(\Phi_{1}^{\ast}(x)\Phi_{2}(x)\Phi^{\ast}_{3}(x)+\mbox{c.c.}) \label{VPhi1} \\
&&+\lambda_{1}|\Phi_{1}(x)|^4+\lambda_{2}|\Phi_{2}(x)|^4 
+\lambda_{3}|\Phi_{3}(x)|^4. \nonumber
\end{eqnarray}

We first substitute 
$$
\Phi_{1}(x)=\sqrt{n_{0}},\, \Phi_{2}(x)=\sqrt{n_{1}},\, 
\Phi_{3}(x)=\sqrt{n_{1}}
$$
into $V(\Phi_{\sigma})$,
\begin{eqnarray}
V(n_{0},n_{1})&=&\lambda_{1}n_{0}^{2}+2\lambda_{2}n_{1}^{2}+
\lambda_{4}n_{0}+2\lambda_{5}n_{1} \nonumber \\
&&-2g\sqrt{n_{0}}n_{1},
\end{eqnarray}
and derive the equations that determine the value of $n_0,\ n_1$ as,
\begin{eqnarray}
\frac{\partial V}{\partial n_{0}}
&=& 2\lambda_{1}n_{0}+\lambda_{4}-g\frac{n_{1}}{\sqrt{n_{0}}}=0,
\nonumber \\
\frac{\partial V}{\partial n_{1}}
&=& 4\lambda_{2}n_{1}+2\lambda_{5}-2g\sqrt{n_{0}}=0,
\label{n0n1}
\end{eqnarray}
where
\begin{eqnarray}
\lambda_{4}=V_{0}-2za_{s}, \\
\lambda_{5}=V_{0}-2za_{h}.
\end{eqnarray}

Next, we introduce three phase fluctuation fields that represent NG bosons
in the FM+2FS phase with $n_0\neq 0,\ n_1\neq 0$,
\begin{eqnarray}
\Phi_{1}(x) &=&\sqrt{n_{0}}+i\phi(x),  \nonumber \\
\Phi_{2}(x) &=&\sqrt{n_{1}}+ib_{1}(x), \nonumber \\ 
\Phi_{3}(x) &=&\sqrt{n_{1}}+ib_{2}(x).
\label{NGfield}
\end{eqnarray}
By substituting Eq.(\ref{NGfield}) into $V(\Phi_\sigma)$ in Eq.(\ref{VPhi1}),
the quadratic terms of the fluctuating fields are obtained as 
\begin{eqnarray}
V_2 =(\phi,b_{1},b_{2})
\left( \begin{array}{ccc}
      2\lambda_{1}n_{0}+\lambda_{4} & -g\sqrt{n_{1}} & g\sqrt{n_{1}}\\
      -g\sqrt{n_{1}} & 2\lambda_{2}n_{1}+\lambda_{5} &-g\sqrt{n_{0}}   \\
       g\sqrt{n_{1}}     & -g\sqrt{n_{0}}         &  2\lambda_{2}n_{1}+\lambda_{5} 
\end{array} \right) \nonumber \\
\times
\left( \begin{array}{c}
\phi \\
b_{1}\\
b_{2}
\end{array} \right). \nonumber
\end{eqnarray}
By using Eq.(\ref{n0n1}), 
\begin{eqnarray}
V_2&=&(\phi,b_{1},b_{2})
 \begin{pmatrix}
      g\frac{n_{1}}{\sqrt{n_{0}}} & -g\sqrt{n_{1}} & g\sqrt{n_{1}}\\
      -g\sqrt{n_{1}}  & g\sqrt{n_{0}} &-g\sqrt{n_{0}}   \\
       g\sqrt{n_{1}}     & -g\sqrt{n_{0}}         &   g\sqrt{n_{0}}
\end{pmatrix}
\left( \begin{array}{c}
\phi \\
b_{1}\\
b_{2}
\end{array} \right)\nonumber\\
&=&(\phi,b_{1},b_{2})g\bf{K}
\left( \begin{array}{c}
\phi \\
b_{1}\\
b_{2}
\end{array} \right).
\end{eqnarray}
It is straightforward to diagonalize the matrix ${\bf K}$ by using a unitary matrix $U$
and obtain the mass gap,
\begin{eqnarray}
U^{-1}{\bf K}U=\begin{pmatrix}
      0 & 0 & 0\\
      0  & 0 & 0   \\
      0   &  0 &   \frac{2n_{0}+n_{1}}{\sqrt{n_{0}}}
\end{pmatrix}.
\end{eqnarray}
Therefore there are two gapless modes that correspond to the NG bosons,
though one may expect three NG bosons as the U(1) pseudo-spin symmetry
is spontaneously broken and also both the $a$ and $b$-atoms Bose condense.
From the above derivation of the mass gap, it is obvious that the
cubic coupling $g\Phi^\ast_1\Phi_2\Phi^\ast_3$ plays an essentially 
important role.

The above result can be understood by returning to the original B-t-J 
Hamiltonian $H_{\rm tJ}$ in Eq.(\ref{HtJ}), and noticing the following
correspondence, 
\begin{equation}
\Phi_1 \sim S^x-iS^y,  \;
\Phi_2 \sim a,  \;
\Phi_3 \sim b.
\label{corresp}
\end{equation}
Furthermore the pseudo-spin $\vec{S}$ is composed of 
$a$ and $b$-atoms like $\vec{S}_i={1 \over 2}B^\dagger_i\vec{\sigma}B_i$ with
$B_i=(a_i,b_i)^t$ in the original B-t-J model.
Therefore a global U(1)$\times$U(1) gauge transformation of $a$ and $b$-atoms
$$
a_i \rightarrow e^{i\theta_a}a_i, \;\;
b_i \rightarrow e^{i\theta_b}b_i,
$$
naturally induces the U(1) transformation of the pseudo-spin,
$$
S^x_i-iS^y_i \rightarrow e^{i(\theta_a-\theta_b)}(S^x_i-iS^y_i). 
$$
This means that the genuine symmetry of the original B-t-J model 
is not U(1)$\times$U(1)$\times$U(1) but U(1)$\times$U(1).
The appearance of 
the term $g\Phi^\ast_1\Phi_2\Phi^\ast_3$ in the effective field theory
eloquently shows this fact.
Then the maximum number of the NG bosons is not three but two, as we 
explicitly demonstrated in the above calculation.

%%%%%%%%%%%%%%%%%%%%%%%%%%%
\subsection{Phase diagram and low-energy excitations: Triangular lattice}

In this subsection we shall focus on the system on the triangular
lattice and study the phase structure and low-energy excitations.
The obtained results will be compared with those by the numerical
study on the quantum XY model given in the previous section and those of the
3D B-t-J model in the following section.
To study the system, we assume a homogeneous state and put 
 $\rho_{1i}=\rho_{2i}=\rho \ (\rho_{3i}=1-\rho_{1i}-\rho_{2i})$.
The field $\Phi_1$ represents pseudo-spin degrees of freedom,
and it is plausible to assume that its condensation has a uniform
amplitude $|\Phi_1|=$constant as holes are distributed homogeneously.
On the other hand, its phase degrees of freedom has a nontrivial
behavior as a result of the frustration.
More general cases will be studied in a future publication.

We first study the effective action by a mean-field theory like
approximation assuming the $\sqrt{3}\times \sqrt{3}$ symmetry.
The triangular lattice is divided into three sublattice labeled $A, B$
and $C$ sublattices.
Each field on the sublattice $A$ is denoted as $\Phi_{\sigma A}$ etc.
It is not so difficult to search the ground state of the potential.
For small $a_{\rm h}$ (i.e., small $t$), $\Phi_2$ and $\Phi_3$ do not
Bose condensate.
On the other hand, a nontrivial long-range oder
of the pseudo-spin appears for $a_{\rm s}>V_0$. 
For constant $|\Phi_1|$, one can show that the state with the three
sublattice coplanar order like
$
\Phi_{1 A}=|\Phi_1|, \ \Phi_{1 B}=|\Phi_1|e^{i2\pi/3}, \
\Phi_{1 C}=|\Phi_1|e^{-i2\pi/3}
$
appears as the lowest-energy state.
This state obviously corresponds to the phase of the $120^o$ spin order 
in Fig.\ref{PD2}.

To obtain the expectation value of $|\Phi_1|$ and to study low-energy excitations,
we relabel the lattice site by dividing it into three sublattices as 
$i\rightarrow (s, o)$ where $o=A, B$ and $C$.
Then we parameterize the field $\Phi_1$ as
\begin{eqnarray}
\Phi_{1A}(s)&=&\rho_i  \nonumber  \\
\Phi_{1B}(s)&=&\rho_i e^{i\frac{2\pi}{3}}  \label{PhiABC} \\
\Phi_{1C}(s)&=&\rho_i e^{-i\frac{2\pi}{3}}. 
\nonumber
\end{eqnarray}
Effective Hamiltonian of the spin part $H^{\rm spin}_{\rm eff}$ is readily obtained 
from $L_{\rm eff}$ in (\ref{Aeff}) (please notice $a_s<0$ in the AF case), 
\begin{eqnarray}
H^{\rm spin}_{\rm eff}&=&\sum_{i,\mu}
\biggl(|a_{s}|\Phi^{\dagger}_{1i}\Phi_{1i+\mu}+\mbox{c.c}\biggr)
\nonumber \\
&&+\sum_{i}\biggl(V_{0}|\Phi_{1i}|^2+\lambda_{3}|\Phi_{1i}|^4\biggr),
\label{Hspin0}
\end{eqnarray}
and substituting Eq.(\ref{PhiABC}) into $H^{\rm spin}_{\rm eff}$, 
\begin{eqnarray}
H^{\rm spin}_{\rm eff}&=&\sum_{i,\mu}\cos{\frac{2\pi}{3}}\biggl(|a_{s}|
\rho_{i}\rho_{i+\mu}+\mbox{c.c}\biggr) \nonumber \\
&&+\sum_{i}\biggl(V_{0}\rho_i^2+\lambda_{1}\rho_i^4\biggr) \nonumber \\
&=& \frac{|a_{s}|}{2}\sum_{i,\mu}|\tilde{\nabla} _{\mu}\rho_{i}|^{2}  \nonumber \\
&&+\sum_{i}\biggl(\lambda_{1}|\rho_{i}|^{4}-(3|a_{s}|-V_{0})|\rho_{i}|^{2}\biggr),
\label{Hspin1}
\end{eqnarray}
where $\tilde{\nabla}_\mu$ denotes the difference operator on the triangular lattice.

From Eq.(\ref{Hspin1}), Bose condensation of $\rho_i$ takes place
for $3a_s>V_0$, and the classical expectation value $\rho_{\rm cl}$ of $\rho_i$
is easily obtained as
\begin{eqnarray}
\rho_{\rm cl}=\sqrt{\frac{3|a_{s}|-V_{0}}{2\lambda_{1}}}.
\end{eqnarray}

\begin{figure}[t]
\begin{center}
\includegraphics[width=5cm]{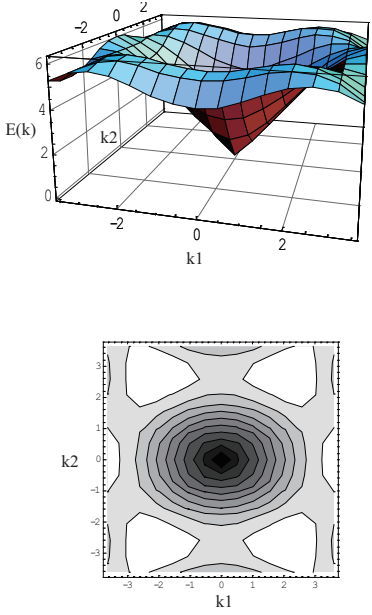}
\vspace{-0.3cm}
\caption{(Color online)
Dispersion relation $E({\bf k})$ in Eq.(\ref{Heff2}).
Parameters are 
$a_{s}=1,
V_{0}=1,
\gamma=5$ and $\beta=1.$
}\vspace{-0.5cm}
\label{dispersion}
\end{center}
\end{figure}

To study the low-energy excitations, we introduce a complex scalar field $\eta_i$
as follows,
\begin{eqnarray}
\rho_{i}=\rho_{\rm cl}+\eta_{i},
\label{eta}
\end{eqnarray}
and then from Eq.(\ref{Hspin0}), we obtain the quadratic terms of $\eta_i$ in 
the Hamiltonian as follows
\begin{eqnarray}
H^{\rm spin}_{\rm eff {\it (2)}}\sim \sum_{\bf{k}}\biggl[\chi(\bf{k})\eta^{\dagger}_{\bf{k}}\eta_{\bf{k}}
+\beta(\eta^{\dagger}_{\bf{k}}\eta^{\dagger}_{-\bf{k}}
+\eta_{\bf{k}}\eta_{-\bf{k}})\biggr]
\label{Hspin2}
\end{eqnarray}
where
\begin{eqnarray}
\chi(\bf{k})&=&-2|a_{s}|\cos{\bigg(\frac{k_{1}a}{2}\biggr)}\cos{\bigg(\frac{\sqrt{3}k_{2}a}{2}\biggr)}  \nonumber \\
&& -|a_{s}|\cos{k_{1}a}+\gamma, \\
\beta&=&\lambda_{1}\rho_{cl}^2,  \; \;
\gamma=4\lambda_{1}\rho_{cl}^{2}+V_{0}.
\end{eqnarray}
We diagonalize Eq.(\ref{Hspin2}) by the Bogoliubov transformation
and obtain
\begin{eqnarray}
H_{\rm eff}=\sum_{{\bf k}}E({\bf k})b^{\dagger}_{\bf{k}}b_{\bf{k}},
\label{Heff1}
\end{eqnarray}
where
\begin{eqnarray}
E({\bf k})=\sqrt{{\chi({\bf k})}^{2}-4\beta^{2}},
\label{Heff2}
\end{eqnarray}
and its typical behavior is shown in Fig.\ref{dispersion}.
As in the limit ${\bf k}\rightarrow 0$,
\begin{eqnarray}
E({\bf k})\rightarrow 0,
\end{eqnarray}
then $b_{\bf{k}}$ represents a NG boson corresponding to the
spontaneous symmetry breaking of the U(1) pseudo-spin symmetry.

%%%%%%%%%%%%%%%%%%%%%%%%%%%%%%%%%%%%%%%%%%%%%%%%%%%%%%%%%%%
%FIG.7
%%%%%%%%%%%%%%%%%%%%%%%%%%%%%%%%%%%%%%%%%%%%%%%%%%%%%%%%%%%

Finally let us study the low-energy excitations in the phase with the $120^o$
spin order {\em plus} 2SF's that corresponds to the phase B in 
Fig.\ref{PD2}.
From Eq.(\ref{Aeff}), the effective Lagrangian $L^{\rm tri}_{\rm eff}$
of low-energy excitations in this phase is obtained as 
\begin{eqnarray}
L^{\rm tri}_{\rm eff}&=&\sum_{i}\biggl(-|a_{s}||\nabla \Phi_{1,i}|^2
+a_{h}|\nabla \Phi_{2,i}|^{2}+a_{h}|
\nabla \Phi_{3,i}|^{2}\biggr)\nonumber\\
&-&\sum_{i}\biggl(\frac{1}{4V_{0}}|\partial_{\tau}\Phi_{1,i}|^{2}+\frac{1}{4V_{0}}|\partial_{\tau}\Phi_{2,i}|^{2}+\frac{1}{4V_{0}}|\partial_{\tau}\Phi_{3,i}|^{2}
\biggr) \nonumber\\
&&+V(\Phi_\sigma),
\label{L'eff}
\end{eqnarray}
\begin{eqnarray}
V(\Phi_{\sigma})&=&\sum_{i}\biggl[\biggr(V_0-2z|a_{s}|\biggr)|\Phi_{1,i}|^2
+\biggr(V_0-2za_{h}\biggr)|\Phi_{2,i}|^2 \nonumber\\
&+&\biggr(V_0-2za_{h}\biggr)|\Phi_{3,i}|^2\biggr)
-g\biggl(\Phi_{1,i}^{\ast}\Phi_{2,i}\Phi^{\ast}_{3,i}+
\mbox{c.c.}\biggr)\nonumber\\
&+&\lambda_{1}|\Phi_{1,i}|^4+\lambda_{2}|\Phi_{2,i}|^4+\lambda_{2}|\Phi_{3,i}|^4\biggr)\biggr],
\label{Veff2}
\end{eqnarray}
where $z$ is again the number of links emanating from a single site.

\begin{figure}[t]
\begin{center}
\includegraphics[width=6cm]{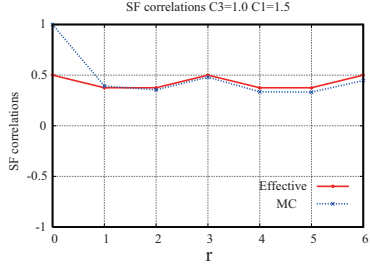}
\vspace{-0.3cm}
\caption{(Color online)
Boson correlation functions $G_{\Phi}(r)=
\langle \Phi^\ast_{2,i}\Phi_{2,i+r} \rangle=
\langle \Phi^\ast_{3,i}\Phi_{3,i+r} \rangle$ in the effective field theory,
and $G_a(r)=G_b(r)$, which is numerically obtained in the quantum XY model.
We set $n_0=n_1=0.5, \ g=6, \ a_{\rm s}=1$ and $a_{\rm h}=1$.
}\vspace{-0.5cm}
\label{corrFs}
\end{center}
\end{figure}

To obtain the ground state, we adopt the following Ansatz for the various condensations,
which respects the three-sublattice symmetry,
\begin{eqnarray}
&&\Phi_{1,A}=\sqrt{n_{0}},
\Phi_{1,B}=\sqrt{n_{0}} e^{i\frac{2\pi}{3}},
\Phi_{1,C}=\sqrt{n_{0}} e^{-i\frac{2\pi}{3}}, \nonumber \\
&&\Phi_{2,A}=\sqrt{n_{1}},
\Phi_{2,B}=\sqrt{n_{1}} e^{i\beta},
\Phi_{2,C}=\sqrt{n_{1}} e^{-i\beta}, \nonumber \\
&&\Phi_{3,A}=\sqrt{n_{1}},
\Phi_{3,B}=\sqrt{n_{1}} e^{-i\beta},
\Phi_{3,C}=\sqrt{n_{1}} e^{i\beta}.
\label{GScondensation}
\end{eqnarray}
The values of $n_0, \ n_1$ and $\beta$ are determined by substituting
Eq.(\ref{GScondensation}) into (\ref{L'eff}) and imposing stationary 
condition,
\begin{eqnarray}
&&{\partial L^{\rm tri}_{\rm eff} \over \partial \beta}
\rightarrow 
12a_{h}n_{1}\sin\beta-4g\sqrt{n_{0}}n_{1}\sin\biggl(\frac{2\pi}{3}-2\beta\biggr)=0,
\nonumber \\
&&{\partial L^{\rm tri}_{\rm eff} \over \partial n_0}
\rightarrow 
2\lambda_{1}n_{0}+\lambda_{4}-g\delta\frac{n_{1}}{\sqrt{n_{0}}}=0, \nonumber \\
&&{\partial L^{\rm tri}_{\rm eff} \over \partial n_1}
\rightarrow 
4\lambda_{2}n_{1}+2\lambda_{5}-2g\delta\sqrt{n_{0}}=0,
\label{gapeq1}
\end{eqnarray}
where $\delta=\frac{1}{3}(1+2\cos(\frac{2\pi}{3}-2\beta))$.
It is verified that nontrivial solutions of $n_0$ and $n_1$ exist
for sufficiently large $a_{\rm h}$ and $a_{\rm s}$
(i.e., negative $\lambda_4$ and $\lambda_5$).
Correlation functions obtained from a typical solution to Eq.(\ref{gapeq1})
are show in Fig.\ref{corrFs},
which have similar behavior to those obtained by the previous MC simulations
of the quantum XY model.

As in the square lattice case, we introduce fields that describe low-energy
excitations,
\begin{eqnarray}
&&\Phi_{1,A}=\sqrt{n_{0}}+i\phi, \nonumber \\
&&\Phi_{1,B}=(\sqrt{n_{0}}+i\phi) e^{i\frac{2\pi}{3}}, \nonumber \\
&&\Phi_{1,C}=(\sqrt{n_{0}}+i\phi) e^{-i\frac{2\pi}{3}},
\label{Phi1}
\end{eqnarray}
\begin{eqnarray}
&&\Phi_{2,A}=\sqrt{n_{1}}+ib_{1}, \nonumber   \\
&&\Phi_{2,B}=(\sqrt{n_{1}}+ib_{1})e^{i\beta}, \nonumber   \\
&&\Phi_{2,C}=(\sqrt{n_{1}}+ib_{1})e^{-i\beta},
\label{Phi2}
\end{eqnarray}
\begin{eqnarray}
&&\Phi_{3,A}=\sqrt{n_{1}}+ib_{2}, \nonumber \\
&&\Phi_{3,B}=(\sqrt{n_{1}}+ib_{2})e^{-i\beta}, \nonumber   \\
&&\Phi_{3,C}=(\sqrt{n_{1}}+ib_{2})e^{i\beta}.
\label{Phi3}
\end{eqnarray}
Substituting the above equations (\ref{Phi1}), (\ref{Phi2}) and (\ref{Phi3})
into $L^{\rm tri}_{\rm eff}$ in (\ref{L'eff}), we obtain the mass term of the
fluctuating fields  as
\begin{eqnarray}
V'_2&=&(\phi,b_{1},b_{2})
 \begin{pmatrix}
      g\delta\frac{n_{1}}{\sqrt{n_{0}}} & -g\delta\sqrt{n_{1}} & g\delta\sqrt{n_{1}}\\
      -g\delta\sqrt{n_{1}}  & g\delta\sqrt{n_{0}} &-g\delta\sqrt{n_{0}}   \\
       g\delta\sqrt{n_{1}}     & -g\delta\sqrt{n_{0}}         &   g\delta\sqrt{n_{0}}
\end{pmatrix}
\left( \begin{array}{c}
\phi \\
b_{1}\\
b_{2}
\end{array} \right)\nonumber\\
&=&(\phi,b_{1},b_{2})g\delta\bf{K}'
\left( \begin{array}{c}
\phi \\
b_{1}\\
b_{2}
\end{array} \right),
\end{eqnarray}
where we have used Eq.(\ref{gapeq1}).
We can easily diagonalize $\bf{K}'$ and get eigenvalues,
\begin{eqnarray}
U^{-1}{\bf K}'U=\begin{pmatrix}
      0 & 0 & 0\\
      0  & 0 & 0   \\
      0   &  0 &   \frac{2n_{0}+n_{1}}{\sqrt{n_{0}}}
\end{pmatrix}.
\end{eqnarray}
The above result indicates that there are only two NG bosons,
though the pseudo-spin U(1) symmetry is spontaneously broken and 
both the $a$ and $b$-atoms Bose condense.

%%%%%%%%%%%%%%%%%%%%%%%%%%%%%%%%%%%%%%%%%%%%%%%%%%%%%%%%%%%
%FIG.8
%%%%%%%%%%%%%%%%%%%%%%%%%%%%%%%%%%%%%%%%%%%%%%%%%%%%%%%%%%%

%%%%%%%%%%%%%%%%%%%%%%%%%%%%
%\bigskip

\section{Phase diagram of the bosonic $\mbox{t-J}$ model on stacked
triangular lattice: Numerical study} 
\setcounter{equation}{0}

In Sec.III.B and IV.C, we studied the low-energy effective theories of the B-t-J model
on the triangular lattice at $T=0$ and obtained phase diagram and 
low-energy excitations.
As we argued previously, if fluctuations of particle density at each site is not
so large, the model at $T=0$ reduces to the system with ``Lorentz invariance", i.e.,
a linear-time derivative term change to a quadratic-time derivative 
term\cite{Lorentz}.
In this case, the imaginary time $\tau$ plays a similar role to
the inter-layer dimension of the stacked 3D lattice.
%%%%%%%%%%%%%%%%%%%%%%%%%%%%%%%%%%%%%%%%%%%%%%%%%%%%%%%%%%%
%FIG.9
\begin{figure}[t]
\begin{center}
\vspace{0.5cm}
\includegraphics[width=6cm]{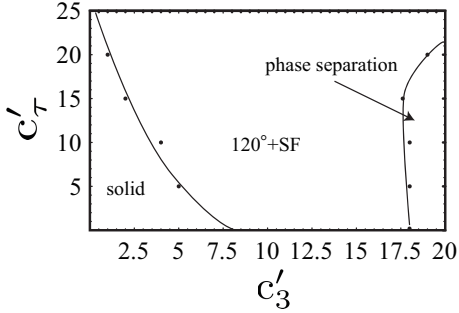}
\vspace{-0.3cm}
\caption{%(Color online)
Phase diagram of the bosonic t-J model in a stacked triangular
lattice. $\rho_a=\rho_b=0.3$, and $C'_1=10.0$.
System size $L=18$.
}\vspace{-0.5cm}
\label{PD3}
\end{center}
\end{figure}
%%%%%%%%%%%%%%%%%%%%%%%%%%%%%%%%%%%%%%%%%%%%%%%%%%%%%%%%%%%
%%%%%%%%%%%%%%%%%%%%%%%%%%%%%%%%%%%%%%%%%%%%%%%%%%%%%%%%%%%
%FIG.10
\begin{figure}[h]
\begin{center}
\vspace{0.5cm}
\includegraphics[width=5cm]{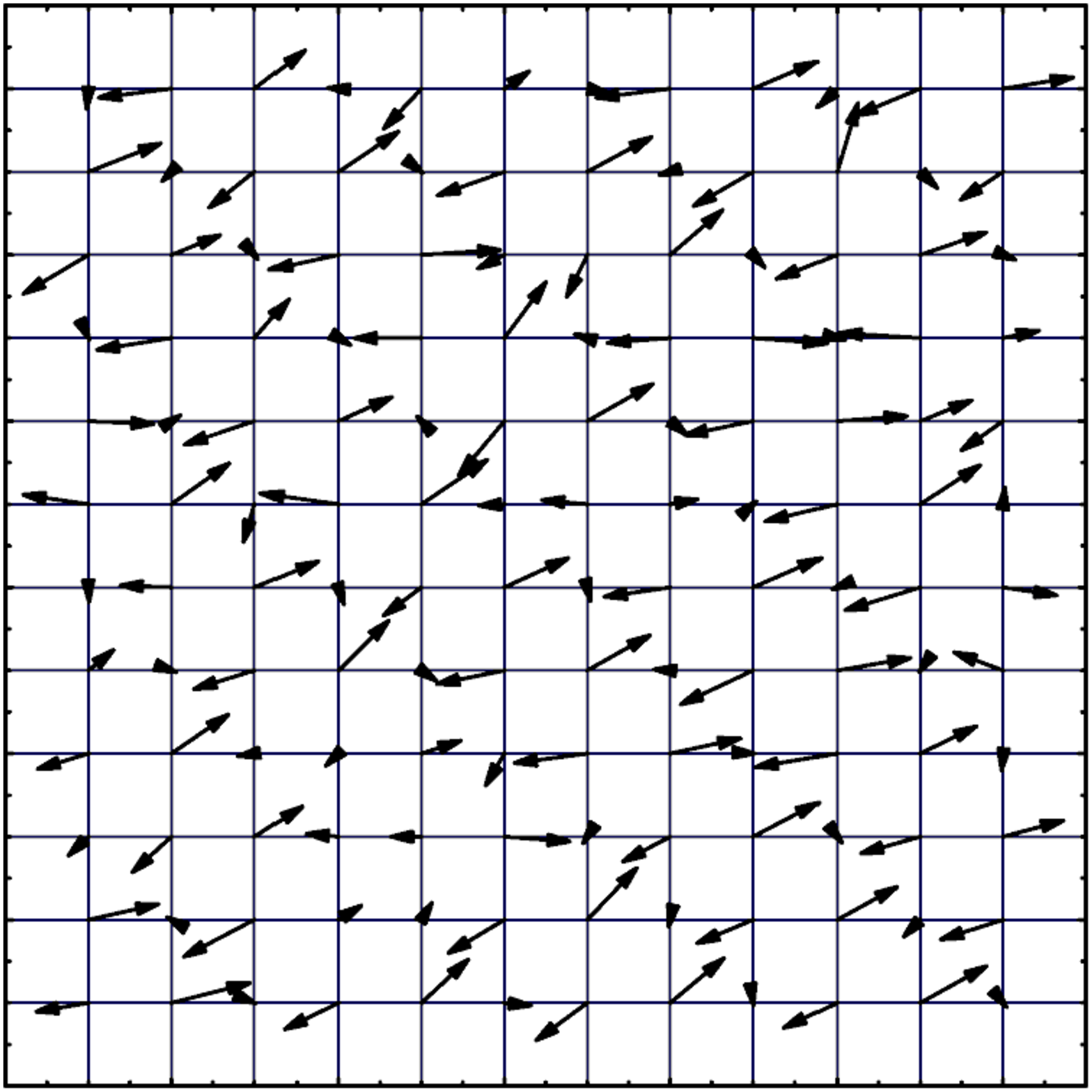} \vspace{1cm} \\
\includegraphics[width=6cm]{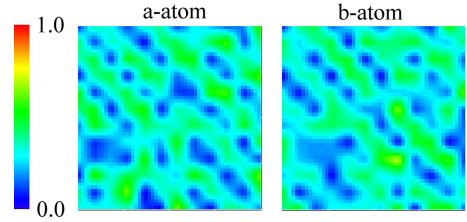}
\vspace{-0.3cm}
\caption{(Color online)
Snapshots of pseudo-spin (upper panel), density of $a$-atom and 
$b$-atom (lower panel).
$S^x-S^y$ is the easy plane of the pseudo-spin (see Fig.\ref{SCBC}),
and length of arrows indicates magnitude of the pseudo-spin.
A stripe oder forms and also there are hole-rich regions.
$L=18$.
}\vspace{-0.5cm}
\label{solid}
\end{center}
\end{figure}
%%%%%%%%%%%%%%%%%%%%%%%%%%%%%%%%%%%%%%%%%%%%%%%%%%%%%%%%%%%

In the present section, we shall study the B-t-J model on the stacked triangular
lattice by numerical simulations with a quasi-classical approximation.
Hamiltonian on the stacked triangular lattice is given as follows,
\begin{eqnarray}
H_{\rm 3DtJ}&=&-\sum_{\langle r,r'\rangle} t(a^\dagger_r a_{r'}
+b^\dagger_{r}b_{r'}+\mbox{h.c.})  \nonumber  \\
&&-\sum_r t'(a^\dagger_{r}a_{r+\hat{3}}
+b^\dagger_{r}b_{r+\hat{3}}+\mbox{h.c.}) \nonumber  \\
&& +J\sum_{\langle r,r'\rangle}(S^x_{r}S^x_{r'}+S^y_{r}S^y_{r'})
%+J_z\sum_{\langle i,j\rangle}S^z_{i}S^z_j,
\label{3DHtJ}
\end{eqnarray}
where $\langle r,r'\rangle$ denotes the NN site of the 2D triangular
lattice, and $\hat{3}$ is the unit vector in the inter-layer direction.
There are (at least) two ways of the MC simulation, i.e., the grand-canonical
and canonical ensemble.
As in the previous studies in this paper, 
we employ the canonical ensemble with the particle number of each atom fixed.
To impose the local constraint, we employ the slave-particle representation
(\ref{slave}).
Then the partition function $Z$ is given as
\begin{equation}
	Z=\int [D\phi D\varphi_1D\varphi_2]e^{-\beta H_{\rm 3DtJ}}.
\label{ZtJ}
\end{equation}
The path integral in (\ref{ZtJ}) is performed by the MC simulation
with local update keeping each particle number fixed.
We call the calculation (\ref{ZtJ}) the quasi-classical approximation
as we ignore the Berry phase in the action of the path integral.
Some detailed discussion on the validity and physical meanings of this approximation
was given in the previous papers\cite{qclassical}.

%%%%%%%%%%%%%%%%%%%%%%%%%%%%%%%%%%%%%%%%%%%%%%%%%%%%%%%%%%%
%FIG.11
\begin{figure}[t]
\begin{center}
\vspace{0.5cm}
\includegraphics[width=5cm]{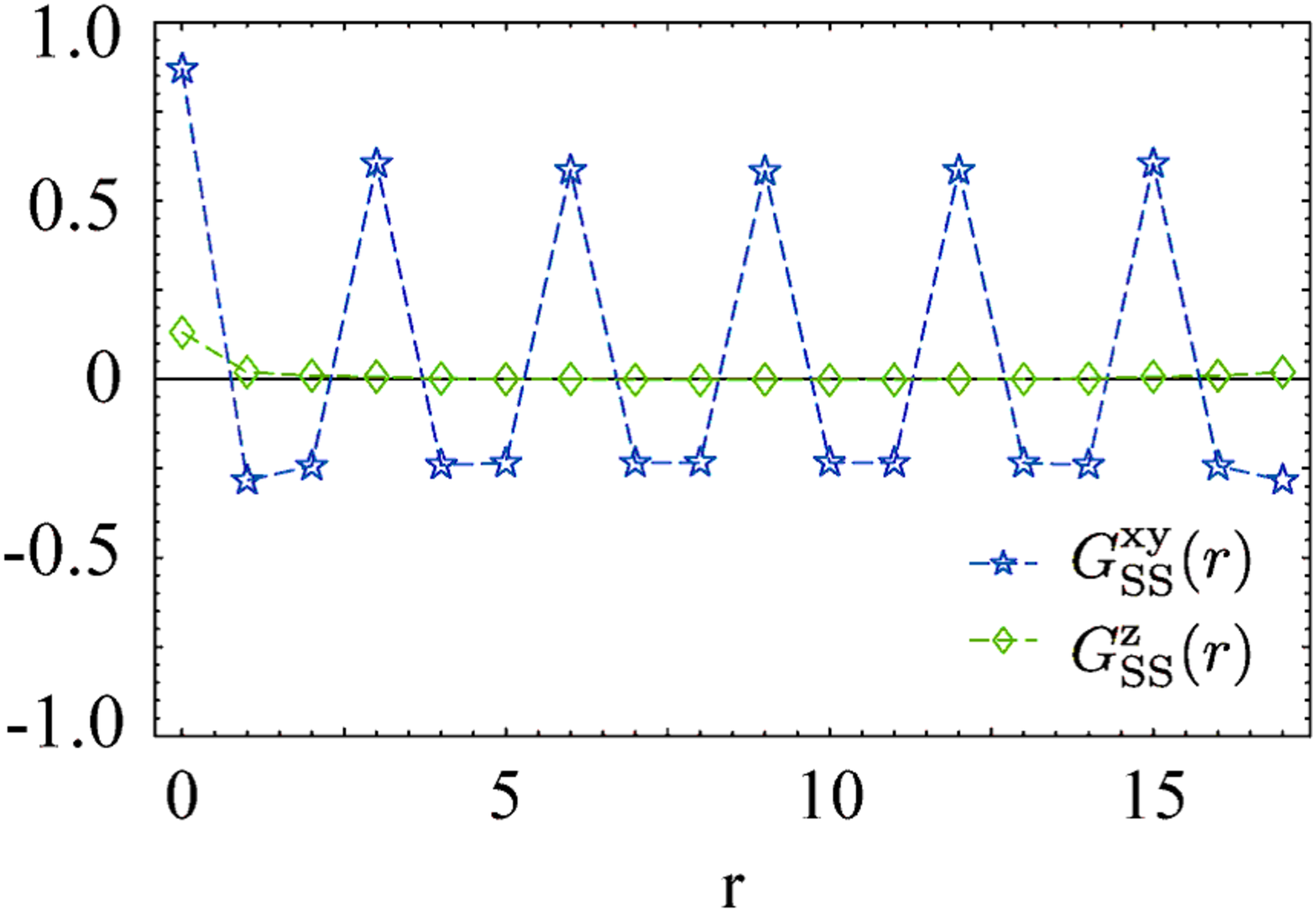} %\vspace{0.5cm} \\
\includegraphics[width=5cm]{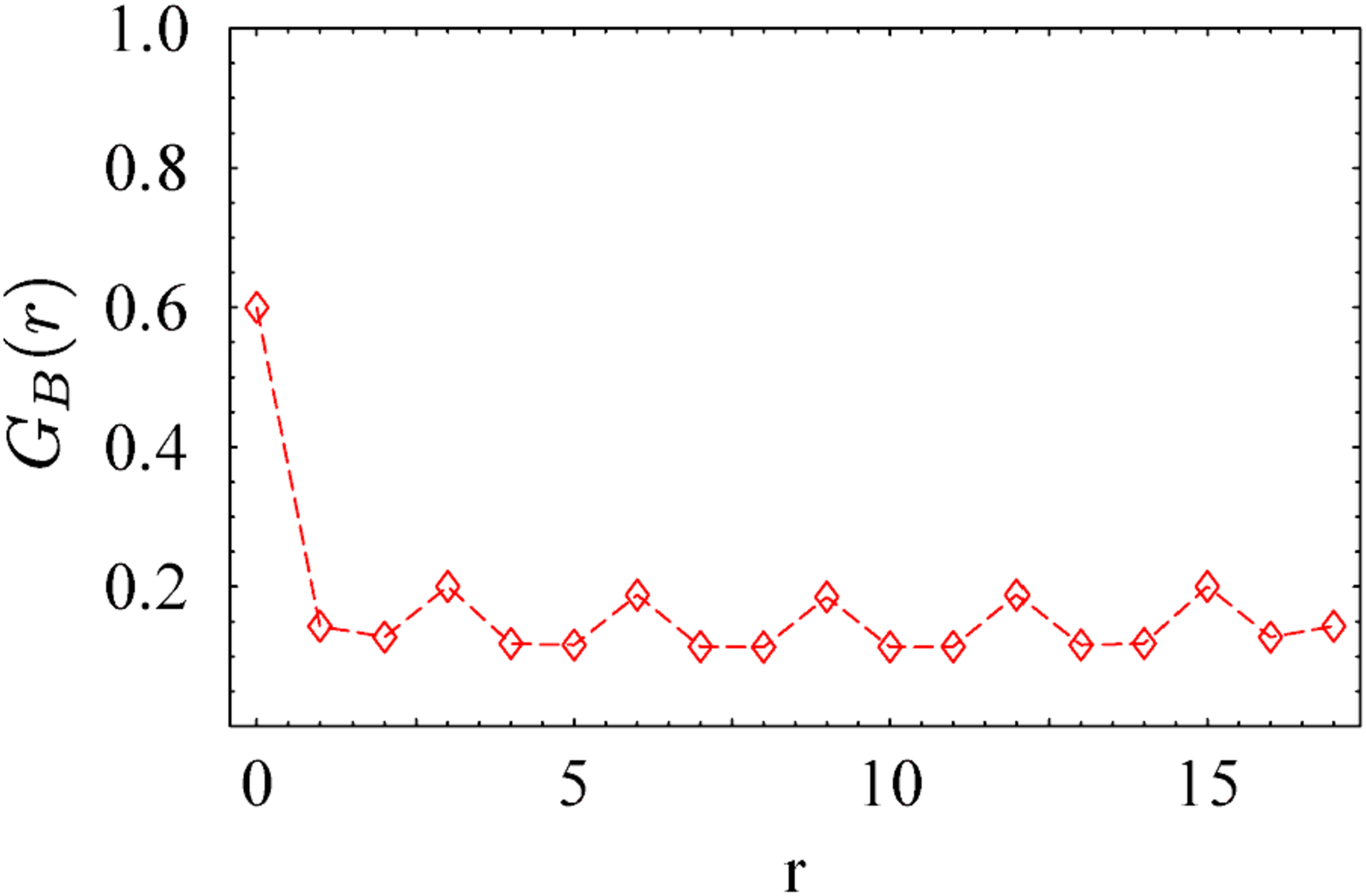}
\vspace{-0.3cm}
\caption{(Color online)
Spin and boson correlation functions obtained by MC simulation 
of the B-t-J model in a stacked triangular lattice.
$C'_\tau=10.0$ and $C'_3=7.0$ in Fig.\ref{PD3}.
$G^{\rm z}_{\rm SS}={4 \over L^3}\sum_i\langle S^z_i S^z_{i+r}\rangle$,
$G^{\rm xy}_{\rm SS}={4 \over L^3}\sum_{i,a=x,y}\langle S^a_i S^a_{i+r}\rangle$,
$G_{\rm B}(r)={1 \over L^3}\sum_i\langle a^\dagger_i a_{i+r} \rangle =
{1 \over L^3}\sum_i\langle b^\dagger_i b_{i+r} \rangle$, where sites $i$ and 
$i+r$ are located in the same 2D triangular lattice.
The result indicates the phase with 120$^o$ spin order and 2SF's.
$L=18$.
}\vspace{-0.5cm}
\label{SCBC}
\end{center}
\end{figure}
%%%%%%%%%%%%%%%%%%%%%%%%%%%%%%%%%%%%%%%%%%%%%%%%%%%%%%%%%%%

We first show the phase diagram for $\rho_a=\rho_b=0.3$ and the hole density $=0.4$
in Fig.\ref{PD3}, where $C'_\tau= t'/(k_{\rm B}T), 
C'_1=J/(k_{\rm B}T)$ and $C'_3=t/(k_{\rm B}T)$ are all dimensionless parameters.
Numerical study was performed for the system size $L=12$ and $18$.
For small hopping amplitude $t$, the system forms a solid with voids whose
snapshot is shown in Fig.\ref{solid}.
As $C'_3$ is increased, phase transition to a state with the 120$^o$ spin order
and 2SF's takes place.
In Fig.\ref{SCBC}, we show the spin and particle correlation functions that verify
this conclusion.
The previous numerical study of the quantum XY model and the
analytical study by the effective field theory predict the existence of
this phase.
This result again indicates a strong resemblance of the phase diagram
of 2D system at $T=0$ and that of the corresponding model in 
stacked 3D lattice.
As $C'_3$ is increased further, phase transition to a phase-separated state
takes place.
In this phase, the system is divided into $a$-atom rich region and $b$-atom
rich region, and in each region a SF of single atom forms.

\vspace{0.5cm}

%%%%%%%%%%%%%%%%%%%%%%%%%%%%
\section{Conclusion}
\setcounter{equation}{0}
In this paper we studied the B-t-J model of the two-component 
bosons with strong on-site repulsions.
We used the salve-particle representation to treat the local constraint faithfully
and derived the effective field theory by integrating out radial degrees of freedom
of the slave bosons.
The resultant field theory is a kind of extended quantum XY model, and
its phase diagram was investigated by the MC simulations.
The results shows that the B-t-J model on the square lattice has a
simple phase diagram whereas that with the AF pseudo-spin coupling
on the triangular lattice has rather complicated structure.

Then we derived the second version of the effective field theory
of the B-t-J model by using a ``Hubbard-Stratonovich transformation"
with the source terms.
The resultant field theory describes the pseudo-spin and Bose condensation
of the $a$ and $b$-atoms directly.
The effective potential of these order parameters clarifies the phase diagram
of the system and verified the existence of the interesting phases like 
$120^0$(pseudo-spin)+2SF's on the triangular lattice.
Furthermore, we studied low-energy excitations, in particular the NG bosons,
in the effective field theory and found that there exist only two NG bosons
even in the phase in which the U(1) symmetry of the pseudo-spin is 
spontaneously broken and the both $a$ and $b$-atoms Bose condense.

Finally we investigated the B-t-J model on the stacked triangular lattice
at finite $T$ by the MC simulations and showed that a similar phase diagram to 
that of  2D system at $T=0$ appears.
This result means that the direction perpendicular to the 2D triangular lattice
plays a similar role to the imaginary-time direction because of the strong
on-site repulsion and the spatial lattice structure.

\bigskip

\acknowledgments 
This work was partially supported by Grant-in-Aid
for Scientific Research from Japan Society for the 
Promotion of Science under Grant No23540301.

%%%%%%%%%%%%%%%%%%%%%%%%%%%%


\begin{references}
\bibitem{HCboson}It is rather straightforward to treat $a_i$ and $b_i$ as 
hard-core bosons as in Refs.\cite{BtJ,BtJ1}.
As we impose the local constraint excluding the doubly-occupied state
in the following calculation, obtained results are qualitatively the same.
\end{references}
\end{document}